%% file: main.tex
\documentclass[twocolumn]{aastex62}

\usepackage{longtable}
\usepackage{graphicx}
\usepackage{amsmath,amssymb}
\usepackage{color}
\usepackage{units}
\usepackage{epstopdf}
\usepackage{hyperref}
\usepackage{multirow}
\usepackage{url}

\usepackage{subfigure}
\usepackage{rotating}

\usepackage[nolist]{acronym}
\input{acronyms}
\usepackage{lineno}
\newcommand{\beq}{\begin{equation}}
\newcommand{\eeq}{\end{equation}}
\newcommand{\bdm}{\begin{displaymath}}
\newcommand{\edm}{\end{displaymath}}

\definecolor{Gray}{gray}{0.9}
\definecolor{orange}{rgb}{0.9,0.5,0}

\newcommand{\ot}{GW170817}

\graphicspath{{./plots/}}

\begin{document}

\title{IN SEARCH OF SHORT GAMMA-RAY BURST OPTICAL COUNTERPARTS WITH THE ZWICKY TRANSIENT FACILITY}

\correspondingauthor{Tom{\'a}s Ahumada}
\email{tahumada@astro.umd.edu}

\author[0000-0002-2184-6430]{Tom{\'a}s Ahumada}\altaffiliation{LSSTC Data Science Fellow}
\affil{Department of Astronomy, University of Maryland, College Park, MD 20742, USA}
\affiliation{Astrophysics Science Division, NASA Goddard Space Flight Center, MC 661, Greenbelt, MD 20771, USA}
\affil{Center for Research and Exploration in Space Science and Technology, NASA Goddard Space Flight Center, Greenbelt, MD 20771, USA}
\author[0000-0003-3768-7515]{Shreya Anand}
\affil{Division of Physics, Mathematics and Astronomy, California Institute of Technology, Pasadena, CA 91125, USA}
\author[0000-0002-8262-2924]{Michael W. Coughlin}
\affil{School of Physics and Astronomy, University of Minnesota, Minneapolis, Minnesota 55455, USA}
\author[0000-0002-8977-1498]{Igor Andreoni}
\affiliation{Joint Space-Science Institute, University of Maryland, College Park, MD 20742, USA}
\affiliation{Astrophysics Science Division, NASA Goddard Space Flight Center, MC 661, Greenbelt, MD 20771, USA}
\affil{Department of Astronomy, University of Maryland, College Park, MD 20742, USA}
\author{Erik C. Kool}	
\affil{The Oskar Klein Centre, Department of Astronomy, Stockholm University, AlbaNova, SE-10691, Stockholm, Sweden}
\author[0000-0003-0871-4641]{Harsh Kumar}	\affil{Department of Physics, Indian Institute of Technology Bombay, Powai, Mumbai 400076, India} \affil{LSSTC Data Science Fellow 2018}
\author{Simeon Reusch} \affil{Deutsches Elektronen Synchrotron DESY, Platanenallee 6, D-15738 Zeuthen, Germany} \affil{Institut f\"ur Physik, Humboldt-Universit\"at zu Berlin, D-12489 Berlin, Germany}
\author[0000-0002-3498-2167]{Ana Sagués-Carracedo}
\affil{The Oskar Klein Centre, Department of Physics, Stockholm University, AlbaNova, SE-106 91 Stockholm, Sweden}
\author{Robert Stein}
\affil{Deutsches Elektronen Synchrotron DESY, Platanenallee 6, D-15738 Zeuthen, Germany}
\affil{Institut f\"ur Physik, Humboldt-Universit\"at zu Berlin, D-12489 Berlin, Germany}
\affil{Division of Physics, Mathematics and Astronomy, California Institute of Technology, Pasadena, CA 91125, USA}
\author{S. Bradley Cenko}
\affiliation{Astrophysics Science Division, NASA Goddard Space Flight Center, MC 661, Greenbelt, MD 20771, USA}
\affiliation{Joint Space-Science Institute, University of Maryland, College Park, MD 20742, USA}
\author[0000-0002-5619-4938]{Mansi M. Kasliwal}
\affil{Division of Physics, Mathematics and Astronomy, California Institute of Technology, Pasadena, CA 91125, USA}
\author[0000-0001-9898-5597]{Leo P. Singer}
\affiliation{Astrophysics Science Division, NASA Goddard Space Flight Center, MC 661, Greenbelt, MD 20771, USA}
\affiliation{Joint Space-Science Institute, University of Maryland, College Park, MD 20742, USA}
% \affil{Department of Astronomy, University of Maryland, College Park, MD 20742, USA}
%
\author{Rachel Dunwoody}\affil{School of Physics, University College Dublin, Dublin 4, Ireland}
\author{Joseph Mangan}\affil{School of Physics, University College Dublin, Dublin 4, Ireland}
\author{Varun Bhalerao}	\affil{Department of Physics, Indian Institute of Technology Bombay, Powai, Mumbai 400076, India}
\author[0000-0002-8255-5127]{Mattia Bulla}
\affiliation{The Oskar Klein Centre, Department of Astronomy, Stockholm University, AlbaNova, SE-10691 Stockholm, Sweden}
\author{Eric Burns} \affil{Department of Physics and Astronomy, Louisiana State University, Baton Rouge, LA 70803, USA}
\author[0000-0002-3168-0139]{Matthew J. Graham}
\affiliation{Division of Physics, Mathematics and Astronomy, California Institute of Technology, Pasadena, CA 91125, USA}
\author[0000-0001-6295-2881]{David L.\ Kaplan}
\affiliation{Center for Gravitation, Cosmology and Astrophysics, Department of Physics, University of Wisconsin--Milwaukee, P.O.\ Box 413, Milwaukee, WI 53201, USA}
\author{Daniel Perley}	\affiliation{Astrophysics Research Institute, Liverpool John Moores University, IC2, Liverpool Science Park, 146 Brownlow Hill, Liverpool L3 5RF, UK}
\author[0000-0002-4694-7123]{Mouza Almualla}
\affil{American University of Sharjah, Physics Department, PO Box 26666, Sharjah, UAE}
\author[0000-0002-7777-216X]{Joshua S. Bloom}
\affiliation{Department of Astronomy, University of California,
Berkeley, CA 94720-3411, USA}
\affiliation{Lawrence Berkeley National Laboratory, 1 Cyclotron Road,
MS 50B-4206, Berkeley, CA 94720-3411, USA}
\author{Virginia Cunningham}
\affil{Space Telescope Science Institute, 3700 San Martin Dr., Baltimore, MD 21218, USA}
\author[0000-0002-0786-7307]{Kishalay De}
\affiliation{MIT-Kavli Institute for Astrophysics and Space Research, 77 Massachusetts Ave., Cambridge, MA 02139, USA}
\author{Pradip Gatkine} \affil{Division of Physics, Mathematics and Astronomy, California Institute of Technology, Pasadena, CA 91125, USA}
\author[0000-0002-9017-3567]{Anna Y. Q.~Ho}
\affiliation{Miller Institute for Basic Research in Science, 468 Donner Lab, Berkeley, CA 94720, USA}
\affiliation{Department of Astronomy, University of California, Berkeley, Berkeley, CA, 94720, USA}
\affiliation{Lawrence Berkeley National Laboratory, 1 Cyclotron Road, MS 50B-4206, Berkeley, CA 94720, USA}
\author{Viraj Karambelkar} \affil{Division of Physics, Mathematics and Astronomy, California Institute of Technology, Pasadena, CA 91125, USA}
\author{Albert K. H. Kong}	\affil{Institute of Astronomy, National Tsing Hua University, Hsinchu 300044, Taiwan}
\author[0000-0001-6747-8509]{Yuhan Yao} \affil{Division of Physics, Mathematics and Astronomy, California Institute of Technology, Pasadena, CA 91125, USA}
\author[0000-0003-3533-7183]{G.C. Anupama}	\affil{Indian Institute of Astrophysics, II Block Koramangala, Bengaluru 560034, India}
\author[0000-0002-3927-5402]{Sudhanshu Barway}	\affil{Indian Institute of Astrophysics, II Block Koramangala, Bengaluru 560034, India}
\author{Shaon Ghosh} \affil{Montclair State University, 1 Normal Ave, Montclair, NJ 07043} \affil{University of Wisconsin-Milwaukee, Milwaukee, WI 53201, USA}
\author{Ryosuke Itoh} \affil{Bisei Astronomical Observatory, 1723-70 Ookura, Bisei-cho, Ibara, Okayama 714-1411, Japan}
\author{Sheila McBreen}\affil{School of Physics, University College Dublin, Dublin 4, Ireland}
%
% BUILDERS
\author[0000-0001-8018-5348]{Eric C. Bellm} \affiliation{DIRAC Institute, Department of Astronomy, University of Washington, 3910 15th Avenue NE, Seattle, WA 98195, USA}
\author{Christoffer Fremling}
\affil{Division of Physics, Mathematics and Astronomy, California Institute of Technology, Pasadena, CA 91125, USA}
\author[0000-0003-2451-5482]{Russ R. Laher}
\affiliation{IPAC, California Institute of Technology, 1200 E. California
	Blvd, Pasadena, CA 91125, USA}
\author[0000-0003-2242-0244]{Ashish~A.~Mahabal}
\affiliation{Division of Physics, Mathematics and Astronomy, California Institute of Technology, Pasadena, CA 91125, USA}
\affiliation{Center for Data Driven Discovery, California Institute of Technology, Pasadena, CA 91125, USA}
\author{Reed L. Riddle} \affiliation{Caltech Optical Observatories, California Institute of Technology, Pasadena, CA 91125}
\author{Philippe Rosnet} \affiliation{Universit\'e Clermont Auvergne, CNRS/IN2P3, LPC, F-63000 Clermont-Ferrand, France.}
\author[0000-0001-7648-4142]{Ben Rusholme}
	\affiliation{IPAC, California Institute of Technology, 1200 E. California
	Blvd, Pasadena, CA 91125, USA}
\author[0000-0001-7062-9726]{Roger Smith}	 \affiliation{Caltech Optical Observatories, California Institute of Technology, Pasadena, CA 91125} 
\author{Jesper Sollerman}
\affil{The Oskar Klein Centre, Department of Astronomy, Stockholm University, AlbaNova, SE-10691, Stockholm, Sweden}
%
% GBM people 

\author{Elisabetta Bissaldi} \affil{Dipartimento di Fisica “M. Merlin” dell’Universit`a e del Politecnico di Bari, I-70126 Bari, Italy} \affil{Istituto Nazionale di Fisica Nucleare, Sezione di Bari, I-70126 Bari, Italy}
\author{Corinne Fletcher} \affil{Science and Technology Institute, Universities Space Research Association, Huntsville, AL 35805, USA}
\author{Rachel Hamburg}\affil{Space Science Department, University of Alabama in Huntsville, 320 Sparkman Drive, Huntsville, AL 35899, USA}
\author{Bagrat Mailyan}\affil{Center for Space Plasma and Aeronomic Research, University of Alabama in Huntsville, 320 Sparkman Drive, Huntsville, AL 35899, USA}
\author{Christian Malacaria}\affil{Astrophysics Office, ST12, NASA/Marshall Space Flight Center, Huntsville, AL 35812, USA}
\author{Oliver Roberts}\affil{Science and Technology Institute, Universities Space Research Association, Huntsville, AL 35805, USA}

\begin{abstract}

The {\it Fermi} \ac{GBM} triggers on-board in response to $\sim$ 40 \acp{SGRB} per year; however, their large localization regions have made the search for optical counterparts a challenging endeavour. We have developed and executed an extensive program with the wide field of view of the \ac{ZTF} camera, mounted on the \ac{P48}, to perform \ac{TOO} observations on 10 {\it Fermi}-\ac{GBM} \acp{SGRB} during 2018 and 2020-2021. Bridging the large sky areas with small field of view optical telescopes in order to track the evolution of potential candidates, we look for the elusive \ac{SGRB} afterglows and \acp{KN} associated with these high-energy events. No counterpart has yet been found, even though more than 10 ground based telescopes, part of the \ac{GROWTH} network, have taken part in these efforts. The candidate selection procedure and the follow-up strategy have shown that \ac{ZTF} is an efficient instrument for searching for poorly localized \acp{SGRB}, retrieving a reasonable number of candidates to follow-up and showing promising capabilities as the community approaches the multi-messenger era. Based on the median limiting magnitude of \ac{ZTF}, our searches would have been able to retrieve a \ot-like event up to $\sim$ 200\, Mpc and SGRB afterglows to z = 0.16 or 0.4, depending on the assumed underlying energy model. Future \acp{TOO} will expand the horizon to z = 0.2 and 0.7 respectively.

\end{abstract}

\section{Introduction}

 Between the years 1969--1972, the Vela Satellites discovered \acp{GRB} and further analysis confirmed their cosmic origin \citep{KlSt1973}. These GRBs are among the brightest events in the universe, and have been observed both in nearby galaxies as well as at cosmological distances \citep{MeDj1997}. The data collected over the years suggest a bimodal distribution in the time duration of the GRB that distinguishes two groups: long GRBs (LGRB\acused{LGRB}; $T_{90}>2s$) and short GRBs (SGRB\acused{SGRB}; $T_{90}<2s$) \citep{KoMe93}, where $T_{90}$ is defined as the duration that encloses the 5th to the 95th percentiles of fluence or counts, depending on the instrument. 

\acp{LGRB} have been associated with \ac{SN} explosions \citep{bloom1999unusual,Woosley2006} and a large number of them have counterparts at longer wavelengths \citep{cano2017}. On the other hand only $\sim35$ \acp{SGRB} have optical/NIR detections \citep{Fong15,rastinejad2021}, thus their progenitors are still an active area of research. \acp{SGRB} have been shown to occur in environments with old populations of stars \citep{berger2005afterglow,davanzo2015} and have long been linked with mergers of compact binaries, such as binary neutron star (BNS) and neutron star–black hole (NSBH) \citep{1992ApJ...395L..83N}. The discovery of the gravitational wave event \ot\, coincident with the short gamma-ray burst GRB 170817A, unambiguously confirmed BNS mergers as at least one of the mechanisms that can produce a \ac{SGRB} \citep{AbEA2017b}. However, compact binary mergers might not be the only source of SGRBs, as collapsars \citep{ahumada2021,Zhang2021} and giant flares from magnetars \citep{burns2021identification} can masquerade as short duration GRBs. Hence, the traditional classification of a burst based solely on the time duration is subject to debate \citep{ZhCh2008,BrNa2013,amati2021}. For example, other gamma-ray properties (i.e. the hardness ratio) can cluster the bursts in different populations \citep{Nakar2007}, and there are a couple of examples for which the time classification of the burst has been questioned due to the presence or lack of \ac{SN} emissions \citep{GalYam2006,ahumada2021,Zhang2021,rossi2021}. In this context, the search for the optical counterparts of SGRBs is essential to unveil the nature of their progenitors and the underlying physics. 

Not all \acp{SGRB} show similar gamma-ray features and different models have tried to explain the observations. For example, the ``fireball'' model \citep{WiRe1997,MeRe1998} describes a highly relativistic jet of charged particle plasma emitted by a compact central engine as a result of a BNS or NSBH merger. The model predicts the production of gamma rays and hard X-rays within the jet. The interaction of the jet and the material surrounding the source produces synchrotron emission in the X-ray, optical, and radio wavelengths. This ``afterglow'' lasts from days to months depending on the frequency range. 

Different models have been applied to the observations that followed GW170817. Among the most popular is the classical case of a narrow and highly relativistic jet powered by a compact central engine \citep{gold17}. Deviations in the light-curves derived from classical models have motivated further developments \citep{WiOb2007,CaGe2009,MeGi2011,DuMa2015}, including Gaussian structured jets \citep{KuGr2003,AbEA2017e,TrPi2017} that can be detected off-axis and do not require the jet to point directly to Earth. Other models predict a more isotropic emission profile, produced by an expanding cocoon formed as the jet makes its way through the ejected material, reaching a Lorentz factor on the order of a few (i.e. $\Gamma \sim $ 2 to 3) \citep{NaHo2014,LaLo2017,KaNa2017,MoNa2017}.

In addition to the GRB afterglow, in the event of a BNS or NSBH merger, the highly neutron rich material undergoes rapid neutron capture ($r$-process), which creates heavy elements and enriches galaxies with rare metals \citep{cote2018origin}. Some of the products of the r-process include radioactive elements; the decay of these newly created elements can energize the ejecta. The produced thermal radiation eventually powers a transient known as a \emph{kilonova} (KN) \citep{LaSc1974,LiPa1998,MeMa2010,Ro2015,KaMe2017}. In the case of an on-axis SGRB, in most cases the optical emission is expected to be dominated by the afterglow and not by the KN.  \citep{gompertz2018diversity,zhu2021kilonova}. There have been attempts to separate the light of the SGRB afterglow and the KN \citep{Fong2016,troja2019afterglow,Ascenzi2019A,oconnor2020tale,Rossi2020kn,Fong2021}, however this still presents a number of challenges. 

Identifying optical counterparts to compact binary mergers can provide a rich scientific output, as demonstrated by the discovery of AT2017gfo \citep{ChBe2017,2017Sci...358.1556C,CoBe2017,2017Sci...358.1570D,2017Sci...358.1565E,KaNa2017,KiFo2017,LiGo2017,2017ApJ...848L..32M,NiBe2017,2017Sci...358.1574S,2017Natur.551...67P,SmCh2017} which led to discoveries in areas as diverse as $r$-process nucleosynthesis, jet physics, host galaxy properties, and even cosmology \citep{KaNa2017,2017Natur.551...64A,2017ApJ...848L..27T,ChBe2017,2017Sci...358.1570D,KaMe2017,2017Natur.551...67P,SmCh2017,TrPi2017}. Previous studies have used the arcminute localizations achieved with the Neil Gehrels {\it Swift} Observatory \ac{BAT} to find and characterize \acp{SGRB} optical counterparts \citep{Fong15,rastinejad2021}, however the number of associations is still only a few dozens. Others have tried following-up thousands of square degrees of \ac{LVC} maps \citep{CoAh19sgrb..131d8001C,Coahu19gw,Andreoni19gw,Goldstein19gw,Andreoni20gw,Hosseinzadeh19gw,Vieira19gw,anand2020nsbh,kasliwal2020kilonova} in the hopes of localizing EM counterparts to gravitational wave events, to no avail. Moreover, other studies have tried to serendipitously find the elusive KN \citep{chatterjee2019toward,andreoni2020constraining,andreoni2021fasttransient}, but they have so far only been able to constrain the local rate of neutron star mergers using wide \ac{FOV} synoptic surveys. 

In this paper we present a summary of the systematic and dedicated optical search of \textit{Fermi}-GBM \acp{SGRB} using the Palomar 48-inch telescope equipped with the 47 square degree \acl{ZTF} camera \citep{graham2019zwicky,Bellm2019zwicky} over the course of $\sim$ 2 years. Previous studies \citep{SiCe2013,SiKa2015} have successfully found optical counterparts to \ac{GBM} LGRBs using the \ac{IPTF} \citep{LaKu2009,RaKu2009}, and  other have serendipitously found orphan afterglows and LGRBs using \ac{ZTF} \citep{andreoni2021fasttransient,2022arXiv220112366H}. There are ongoing projects like Global MASTER-Net \citep{lipunov2005master}, and the Gravitational-Wave Optical Transient Observe (GOTO\acused{GOTO}; \citealt{mong2021searching}) that are using optical telescopes to scan the large regions derived by \ac{GBM}. We note that the optical afterglows of LGRBs are usually brighter than of SGRBs, thus the \ac{TOO} strategy might differ from the one presented in this paper. We base our triggers on \ac{GBM} events since \ac{GBM} is more sensitive to higher energies than \textit{Swift} and it detects SGRBs at four times the rate of \textit{Swift}, making it the most prolific compact binary merger detector. 

In section \ref{sec:ObsNData} we describe the facilities involved along with the observations and data taken during the campaign. We describe our filtering criteria and how candidates are selected and followed up in section \ref{sec:candidates}, and detail the \textit{Fermi} events we followed up in section \ref{sec:sgrb_description}. In section \ref{sec:upperlimits} we compare our observational limits to \ac{SGRB} transients in the literature. In section \ref{sec:non-det} we discuss the implications of the optical non-detection of a source and we explore the sensitivity of our searches. Using the lightcurves of the transients generated for our efficiency analysis, we put the detection of an optical counterpart in context for future \ac{TOO} follow-up efforts in section \ref{sec:follow-up-plan}. We summarize our work in section \ref{sec:conclusion}.

\section{Observations and Data}\label{sec:ObsNData}

In this section we will broadly describe the characteristics of the telescopes and instruments involved in this campaign, as well as the observations. We start with the \emph{Fermi}-\ac{GBM}, our source of compact mergers, followed by \ac{ZTF}, our optical transient discovery engine, and finally describe the facilities used for follow-up. 

\subsection{\textit{Fermi} Gamma-ray Burst Monitor} \label{sec:Fermi}
The Gamma-ray Burst Monitor (GBM) is an instrument on board the \textit{Fermi} Gamma-ray Space Telescope sensitive to gamma-ray photons with energies from 8\, keV to 40\, MeV \citep{Meegan_2009}. The average rest frame energy peak for \acp{SGRB} ($E_{p,i}\sim 0.5$ MeV; \citealt{zhang2012correlation}) is enclosed in the observable \ac{GBM} energy range and not in the \textit{Swift} BAT energy range (5-150 keV). Additionally, any given burst should be seen by a number of detectors, as \ac{GBM} is sensitive to gamma-rays from the entire unocculted sky.

The low local rate of \textit{Swift} SGRBs has impeded the discovery of more \ot-like transients \citep{dichiara2020short}. On the other hand, GBM detects close to 40 \acp{SGRB} per year \citep{Meegan_2009}, four times the rate of \textit{Swift}. However, the localization regions given by \ac{GBM} usually span a large portion of the sky, going from a few hundred sq. degrees to even a few thousand square degrees. These large regions make the systematic search for counterparts technically challenging and time consuming \citep{fermi2020cat,goldstein2020evaluation}.

Our adopted strategy prioritizes \textit{Fermi}-\ac{GBM} \ac{GRB} events visible from Palomar that present a hard spike, that are classified as \acp{SGRB} by the on-board GBM algorithm, and that are not detected by \textit{Swift}. During the first half of our campaign (2018), we did not have any constraints on the size of the GRB localization region. However, during the second half of our campaign, we restricted our triggers to the events for which more than 75\% of the error region could be covered twice in $\sim$2 hrs. With \ac{ZTF} this corresponds to a requirement that 75\% of the map encloses less than ~500 deg$^2$., which explains the difference in the number of triggers between the first and second half of our campaign.

\begin{figure*}
 \includegraphics[width=1\textwidth]{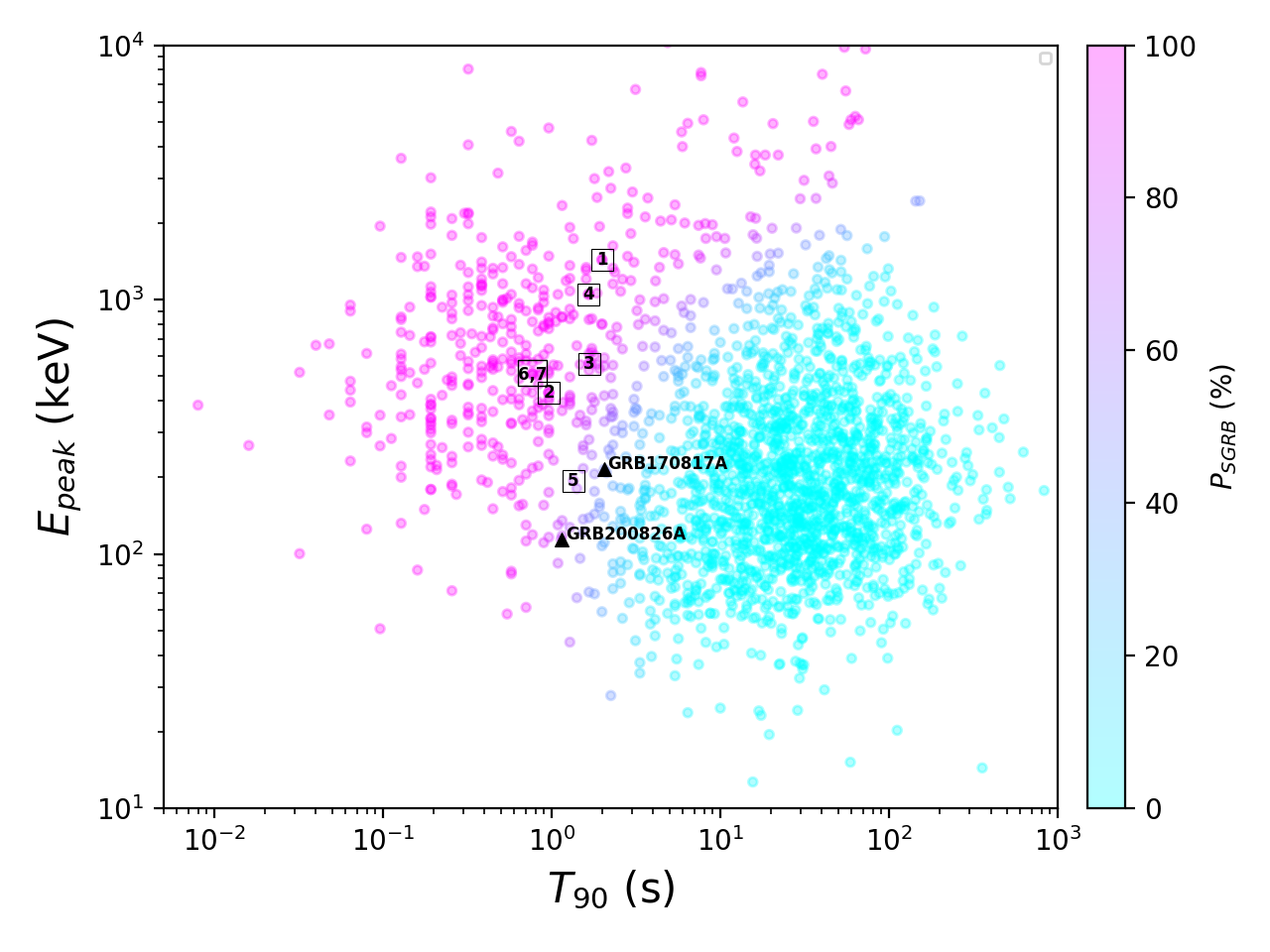}
 \caption{The peak energy based on a Comptonized fit, $E_{peak}$ (keV), versus the time-integrated $T_{90}$ (s), for 2,310 \textit{Fermi} GBM GRBs. The data are fit with two log-normal distributions for the two GRB classes. The colour of the data points indicates the probability, with magenta being 100\% SGRB and cyan being 100\% LGRB. We show in squares numbered from 1 to 7 the following SGRBs: GRB 180523B, GRB 180626C, GRB 180715B, GRB 181126B, GRB 210510A,  GRB 180913A and GRB 180728B. Note that the GRB 180728B and GRB 180913A share the same location in this parameter space. The bursts GRB 200514B and GRB 201130A are not shown as the power-law model is preferred over the Comptonized fit, thus there is no $E_{peak}$ parameter associated to them. For context, we show in triangles GRB 170817A and GRB 200826A.}
 \label{fig:EpT90}
\end{figure*}

For each GRB, we calculate the probability of belonging to the population that clusters the SGRBs based on their comptonized energy peak $E_{peak}$ and their duration $T_{90}$. For this, we fit two log-normal distributions (representing the long and short classes) to a sample of ~ 2300 GRBs. We derive and color code the probability $P_{SGRB}$ by assessing where each GRB falls in the distribution (see Fig. \ref{fig:EpT90}, and \citealt{ahumada2021} for more details). In Table \ref{table:GRBs} we list the relevant features of the \acp{SGRB} selected for follow-up. 

 \begin{deluxetable*}{ccccccccc}
 \tablecaption{ \label{table:GRBs} Global features of the \emph{Fermi}-\ac{GBM} \ac{SGRB} followed-up with ZTF. The peak energies come from the public \textit{Fermi} catalog \citep{fermi2020cat} for GRB 180523B, GRB 180626C, GRB 180715B, GRB 180913A and GRB 181126B. Additionally, we compiled $E_p$ listed in \citet{gcn23057} for GRB 180728B, and independently provide time-integrated fits for GRB 200514B, GRB 201130A, and GRB 210510A over the $T_{90}$. We list the GRB name, their trigger number, the Julian day (JD) of each event, the $T_{90}$ duration, the area encompassed by the 90\% (50\%) credible region (C.R.), the signal-to-noise ratio from the \textit{Fermi} detection, the peak energy of the gamma-ray spectrum ($E_{peak}$), the fluence of the burst, and the probability of the burst to belong to the SGRB population (see Sec. \ref{sec:Fermi}). The area associated to a given C.R. is derived by calculating the number of pixels that cumulatively sum a specific percentage, using the HEALPix map of each GRB. For events with a \dag, the power-law model is preferred over the comptonized model, thus there is no $E_p$ parameter. We show separately the parameters of GRB 200826A, as it was not related to a compact binary merger \citep{ahumada2021}}
\tablehead{\colhead{GRB} & \colhead{Fermi Trigger}&\colhead{{\begin{tabular}{@{}c@{}}Time \\ $[$JD$]$\end{tabular}}}& \colhead{{\begin{tabular}{@{}c@{}}$T_{90}$ \\ $[$s$]$\end{tabular}}} & \colhead{{\begin{tabular}{@{}c@{}}90\% (50\%) C.R. \\ $[deg^2]$\end{tabular}}} &\colhead{S/N} & \colhead{\begin{tabular}{@{}c@{}}$E_{peak}$ \\ $[$keV$]$\end{tabular}}& \colhead{\begin{tabular}{@{}c@{}}Fluence \\ $[10^{-8}$ erg/cm$^2]$\end{tabular}}&\colhead{$P_{SGRB}$}
}
\startdata
GRB 180523B & 548793993 & 2458262.2823 & 2.0 $\pm$ 1.4& 5094 (852) &  6.9 & $1434 \pm 443$ & 25.7$\pm$ 2.3 & 0.99 \\	
GRB 180626C & 551697835 & 2458295.8916 & 1.0 $\pm$ 0.4& 5509 (349) &  7.1 & $431  \pm 81$ & 49.1 $\pm$ 3.8 &  0.97 \\
GRB 180715B & 553369644 & 2458315.2412 & 1.7 $\pm$ 1.4 & 4383 (192) & 12.5 & $560  \pm 89$ & 52.0 $\pm$ 1.7&   0.92\\
GRB 180728B & 554505003 & 2458328.3819 & 0.8 $\pm$ 0.6& 397 (47)&  20.2 & $504  \pm 61$ &  130.9 $\pm$ 2.0 &0.99 \\
GRB 180913A & 558557292 & 2458375.2834 & 0.8 $\pm$ 0.1& 3951 (216) & 10.0 & $508  \pm 90$ & 79.1 $\pm$ 2.0&  0.99\\ 
GRB 181126B & 564897175 & 2458448.6617 & 1.7 $\pm$ 0.5& 3785 (356) &  7.5  & $1049 \pm 241$ & 48.3 $\pm$ 3.2 & 0.99 \\ 
GRB 200514B & 611140062 & 2458983.8802 & 1.7 $\pm$ 0.6& 590 (173) &  5.1  & \dag &17.8 $\pm$ 1.1 &  -- \\
GRB 201130A & 628407054 & 2459183.7297 & 1.3 $\pm$ 0.8& 545 (139) &  5.3  & \dag & 37.0 $\pm$ 5.2 &  -- \\
GRB 210510A & 642367205 & 2459345.3055 & 1.3 $\pm$ 0.8& 1170 (343) &  5.6  & $194 \pm 60$ & 23.2 $\pm$ 1.4 & 0.74\\
\hline
GRB 200826A & 620108997 & 2459087.6874 & 1.1 $\pm$ 0.1& 339 (63) &  8.1 & $88.9 \pm 3.2$ &426.5 $\pm$ 2.2&  0.74\\
\hline
\enddata
\end{deluxetable*}

\subsection{The Zwicky Transient Facility}

We have used \ac{ZTF} to scan the localization regions derived by the \emph{Fermi}-\ac{GBM}. 
\ac{ZTF} is a public-private project in the time domain realm which employs a dedicated camera \citep{2020PASP..132c8001D} on the Palomar 48-inch Schmidt telescope. The \ac{ZTF} field of view is 47 deg$^2$, which usually allows us to observe more than 50\% of the \ac{SGRB} error region in less than one night. The public \ac{ZTF} survey \citep{2019PASP..131f8003B} covers the observable northern sky every two nights in $g$- and $r$-bands with a standard exposure time of 30\,s, reaching an average $5\sigma$ detection limit of $r = 20.6$. 

Two \ac{TOO} strategies were tested during this campaign, one during 2018 and the second during 2020-2021. Most modifications came after lessons learned during the follow-up efforts of gravitational waves in 2019 \citep{Coahu19gw,anand2020nsbh,kasliwal2020kilonova}. The original \ac{TOO} observing plan allowed us to start up to 36 hrs from the SGRB GBM trigger. However, since the afterglow we expect is already faint ($m_r>19$ mag) and fast fading ($\Delta m/ \Delta t > 0.3$ mag per day), our revised strategy only includes triggers that can be observed from Palomar within 12 hrs. The exposure time for each trigger ranges from 60\,s to 300\,s depending on the size of the localization region, as there is a trade-off between exposure time and coverage. We generally prioritized coverage over depth, and for the second half of our campaign, we only triggered on maps where more than 75\% of the region could be covered. The same sequence is repeated a second time the following night, unless additional information from other spacecraft modifies the error region. Generally, fields with an airmass $>$ 2.5 are removed from the observing plan. 
 
 We schedule two to three sets of observations depending on the visibility of the region, using the ZTF $r$- and $g$-bands. The combination of $r$- and $g$-band observations was motivated by the need to look for afterglows and KNe, which are both fast evolving red transients. In fact, the SGRB afterglows in the literature show red colors (i.e. $g-r>0.3$ mag) and a rapid evolution, fading faster than $\Delta m_r/\Delta t > 0.5$ mag per day. On the other hand, \ot\ started off with bluer colors and evolved dramatically fast in the optical during the first days, with $g-r = 0.5$ mag ~1 day after the \emph{Fermi} alert and $\Delta m_g/\Delta t > 1$ mag per day. Even though we expect a fast fading transient, if we assume conservative fading rates of 0.3-0.5 mag per day, we would need observations separated by 8 to 5 hrs respectively to detect the decline using ZTF data with photometric errors of the order of 0.1 mag. This \ac{TOO} strategy thus relies on the color of transients for candidate discrimination, as this is easier to schedule than multi-epoch single-band photometry within the same night and with sufficient spacing between observations.
 
\begin{figure*}[t]
\includegraphics[width=0.49\linewidth]{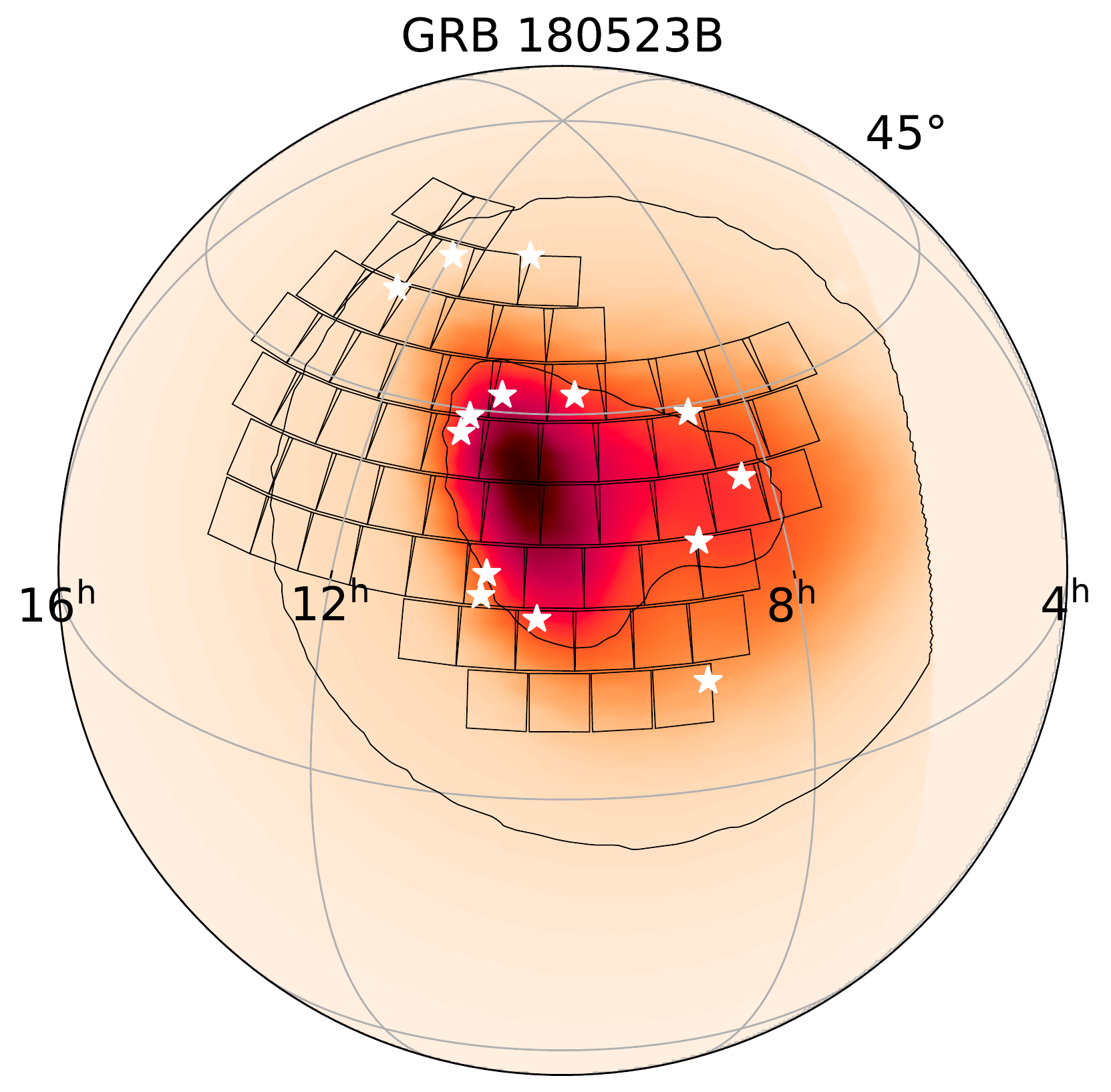}
\includegraphics[width=0.49\linewidth]{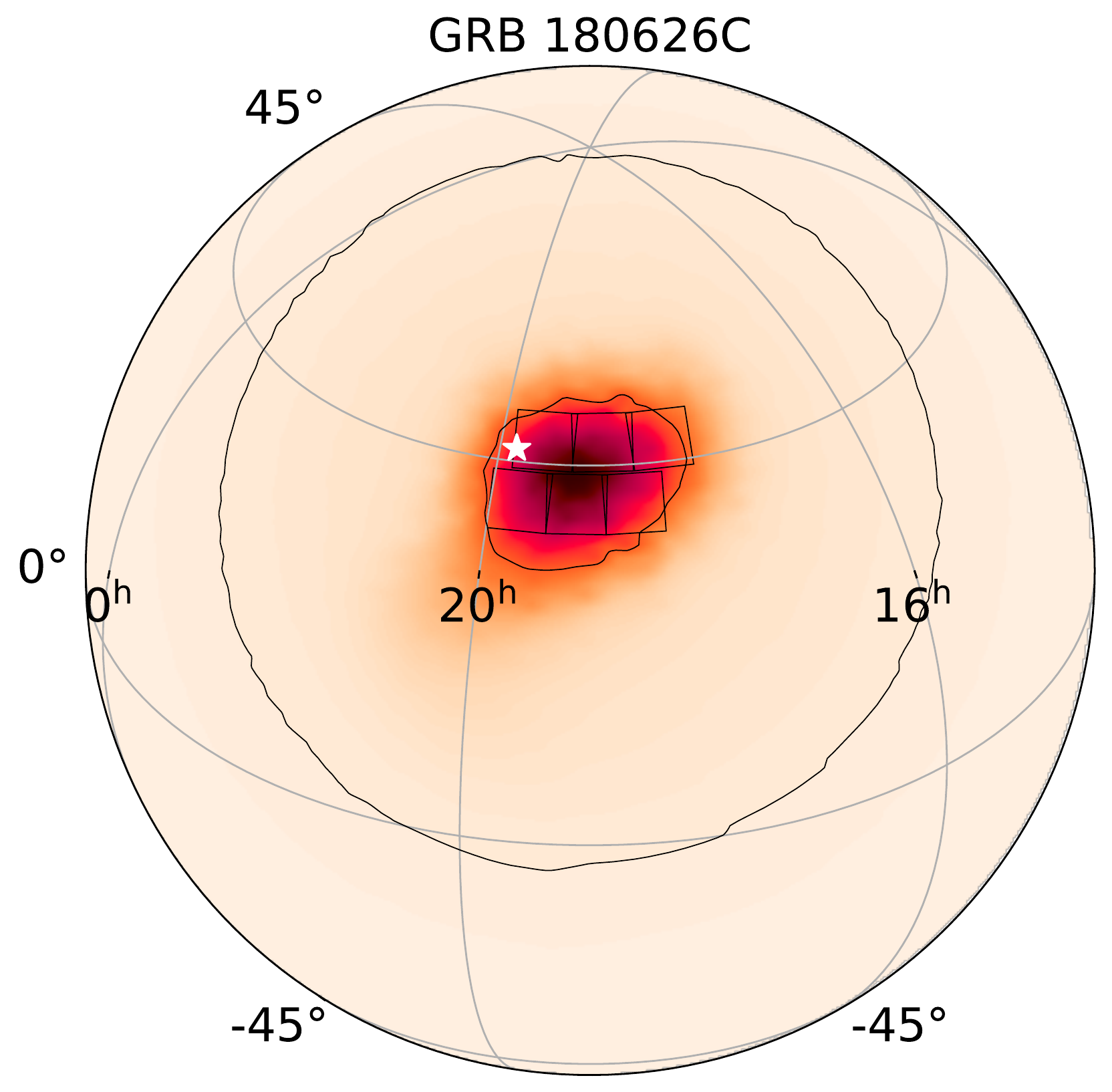}
\includegraphics[width=0.49\linewidth]{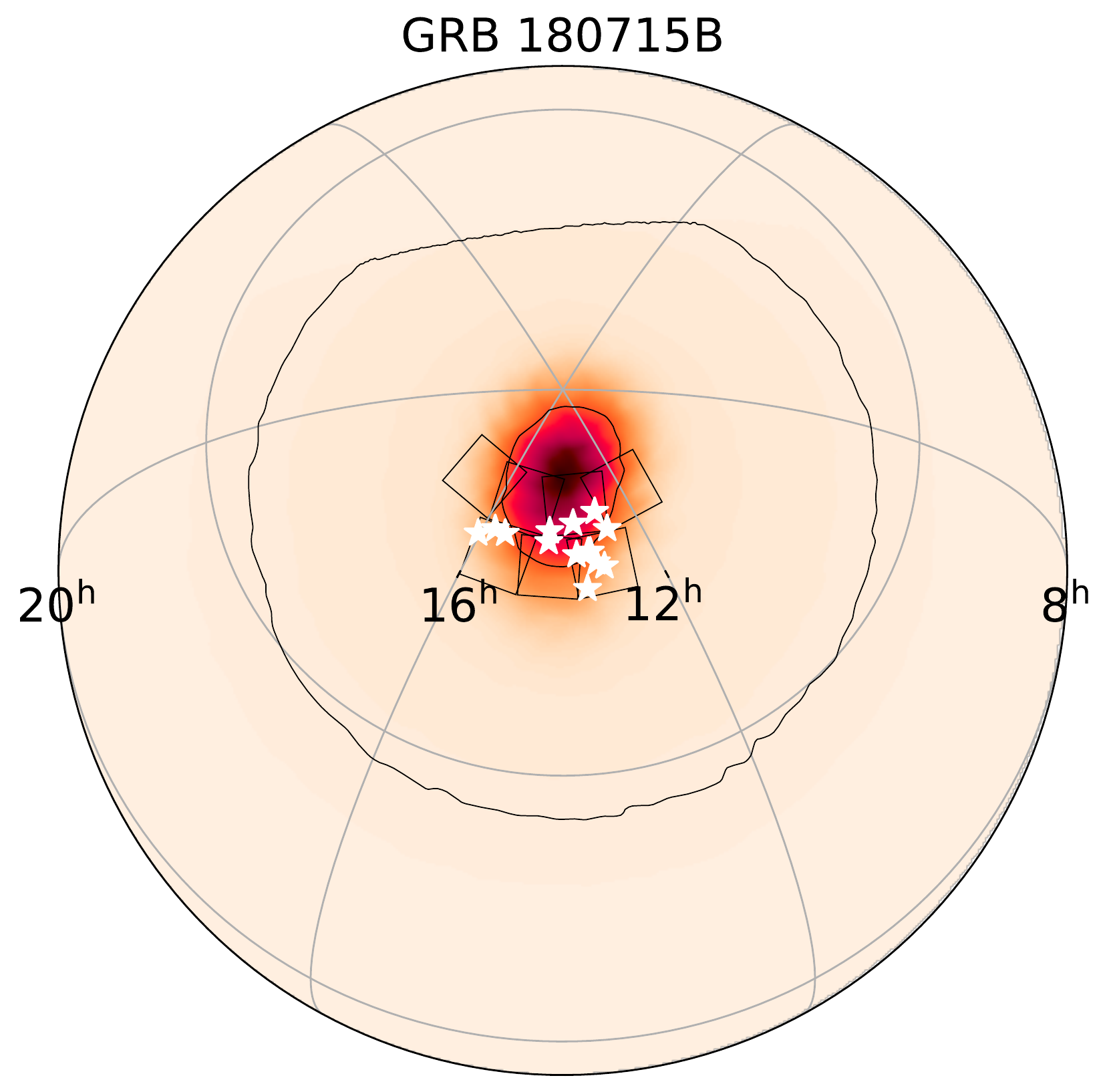}
\includegraphics[width=0.49\linewidth]{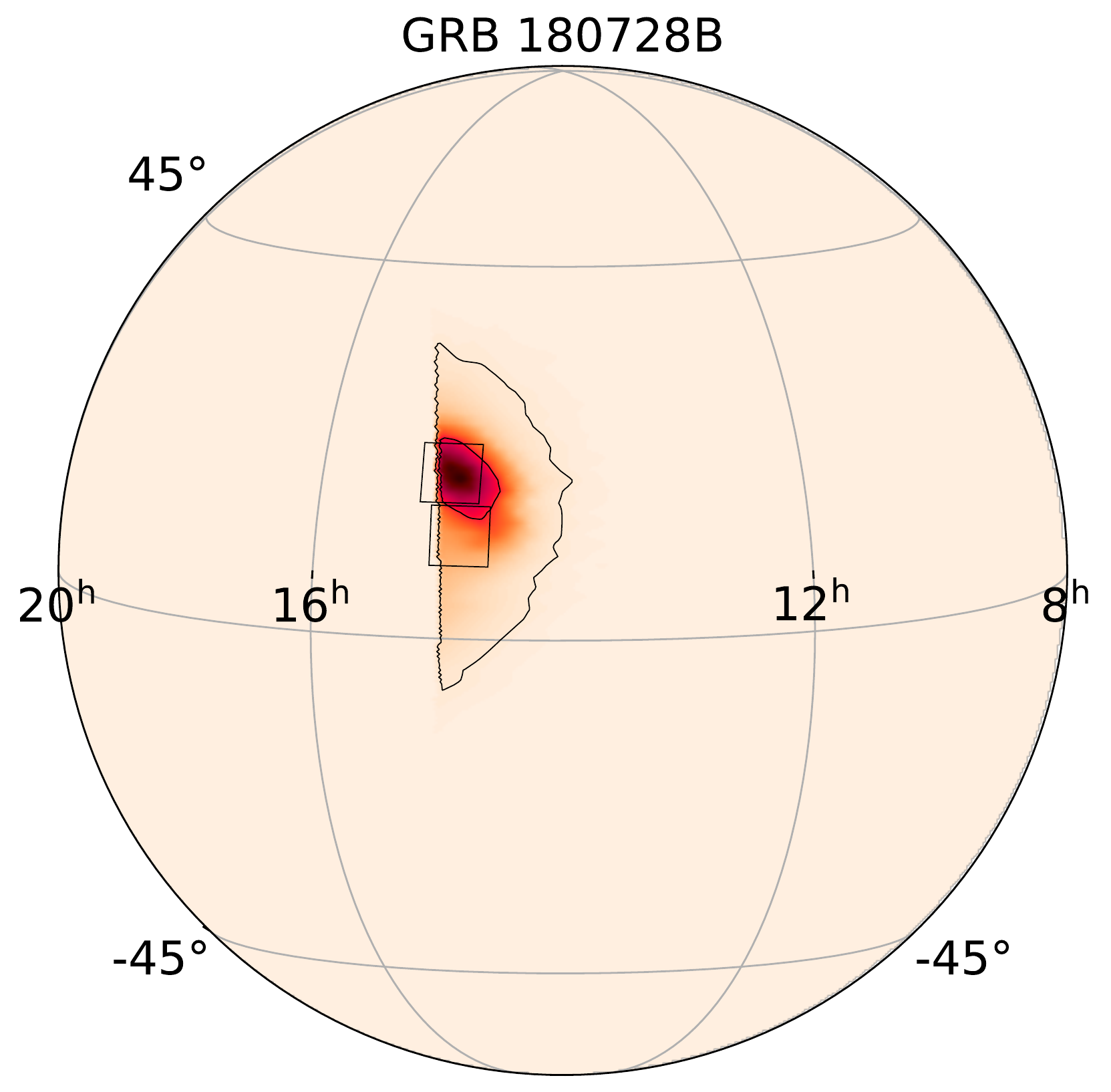}
\caption{ Coverage of four \ac{ZTF} triggers and their \textit{Fermi} \ac{GBM} localization regions. Starting on the top left, the skymaps of GRB 180523, GRB 180626, GRB 180715, and GRB 180728 are shown along the $\approx$ 47 deg$^2$ \ac{ZTF} tiles (black quadrilaterals). The 50\% and 90\% credible regions are shown as black contours and the sources discovered during the \ac{ZTF} trigger as white stars (details in Section~\ref{sec:sgrb_description}). The grid shows the Right Ascension in hours and the Declination in degrees.}
\label{fig:skymap1}
\end{figure*}

\begin{figure*}[t]
\includegraphics[width=0.49\linewidth]{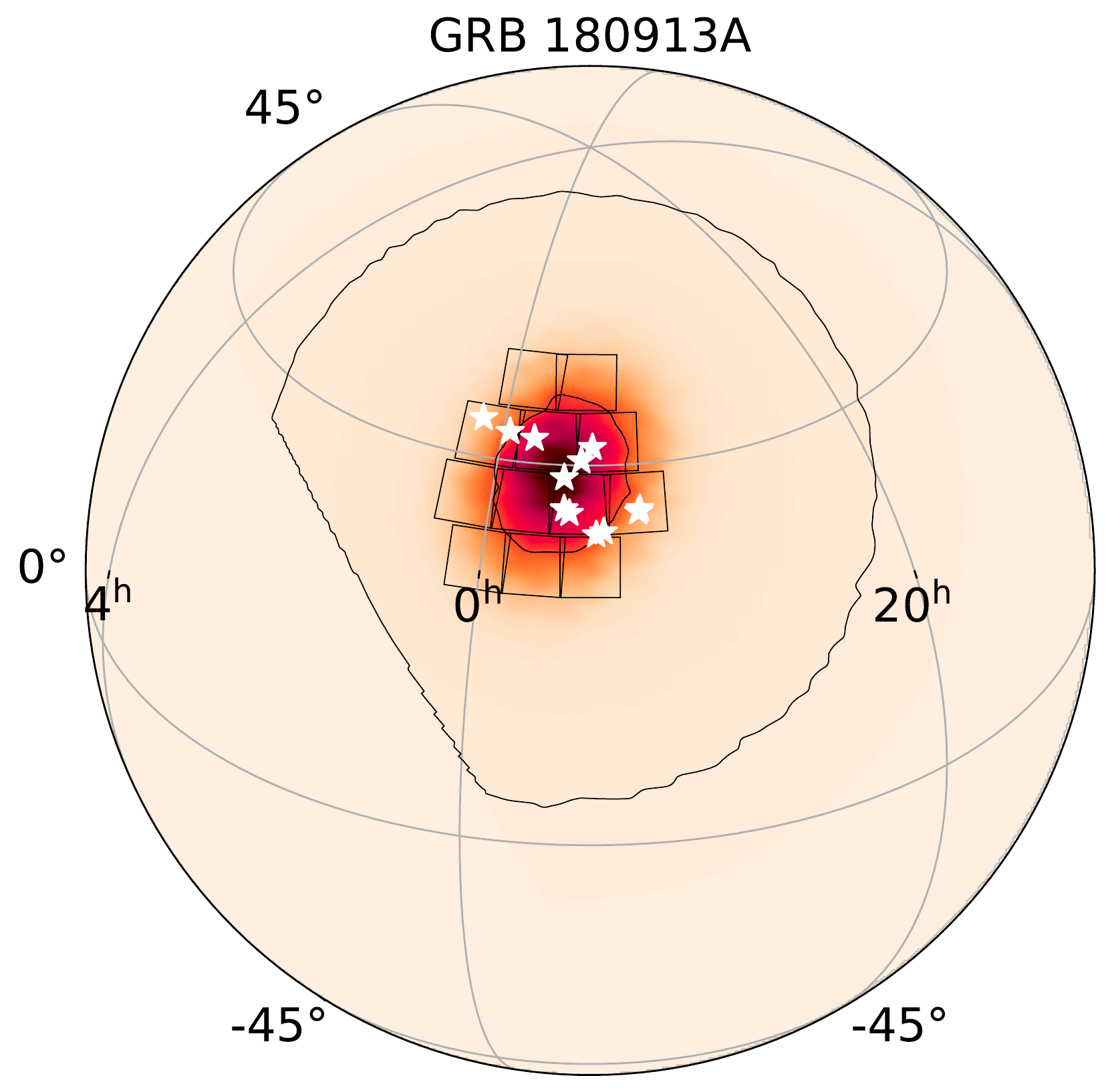}
\includegraphics[width=0.49\linewidth]{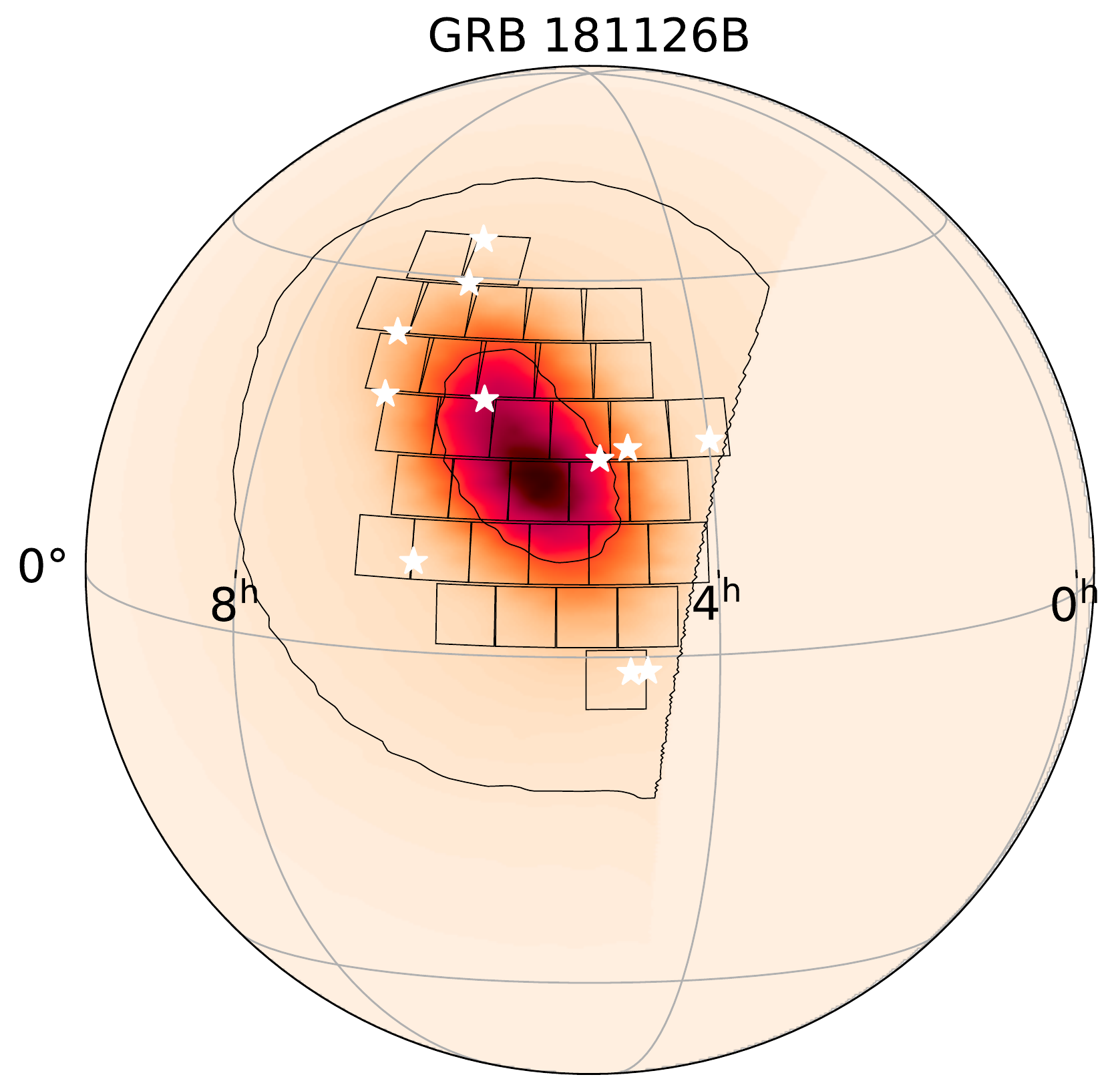}
\includegraphics[width=0.49\linewidth]{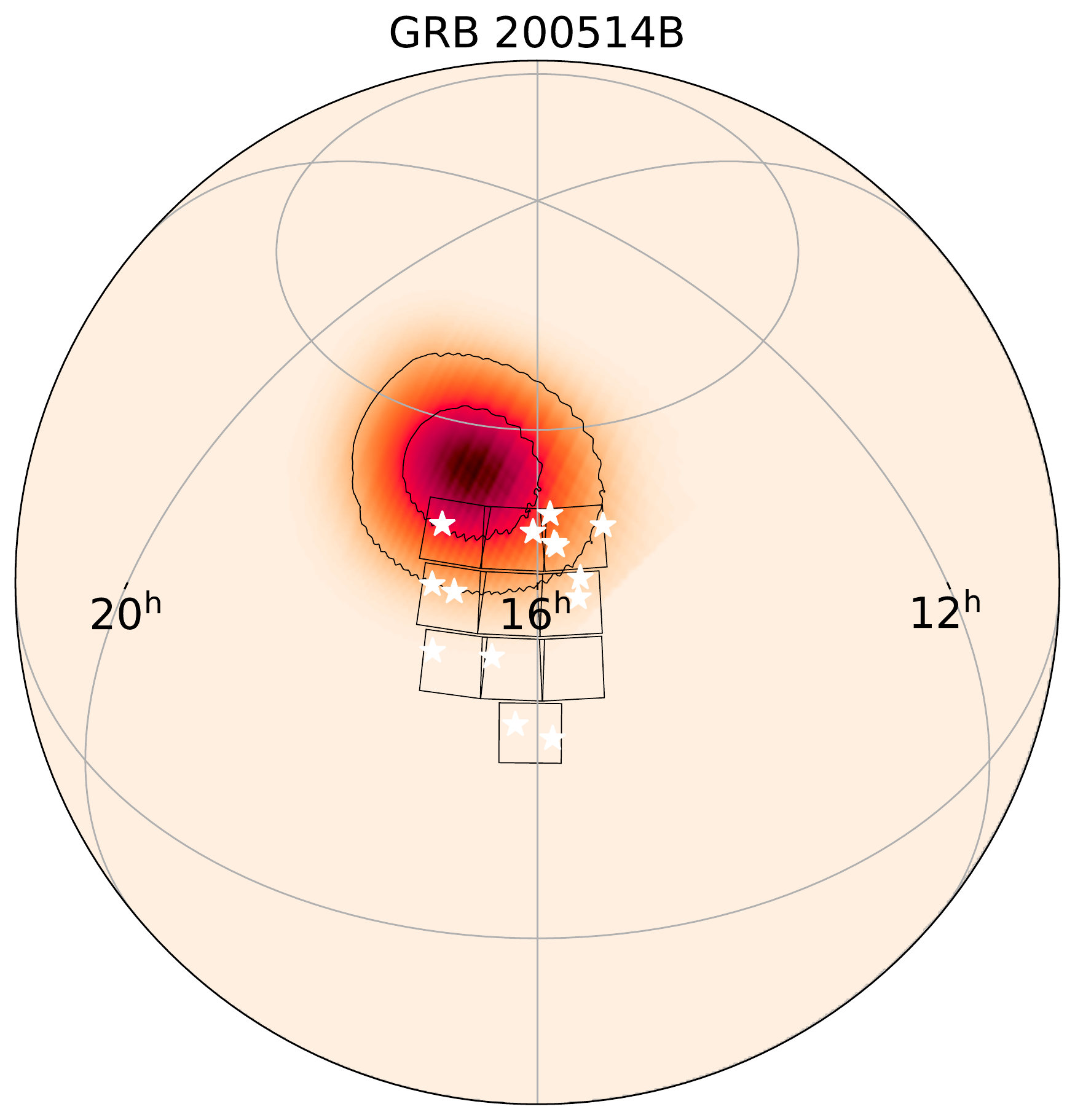}
\includegraphics[width=0.49\linewidth]{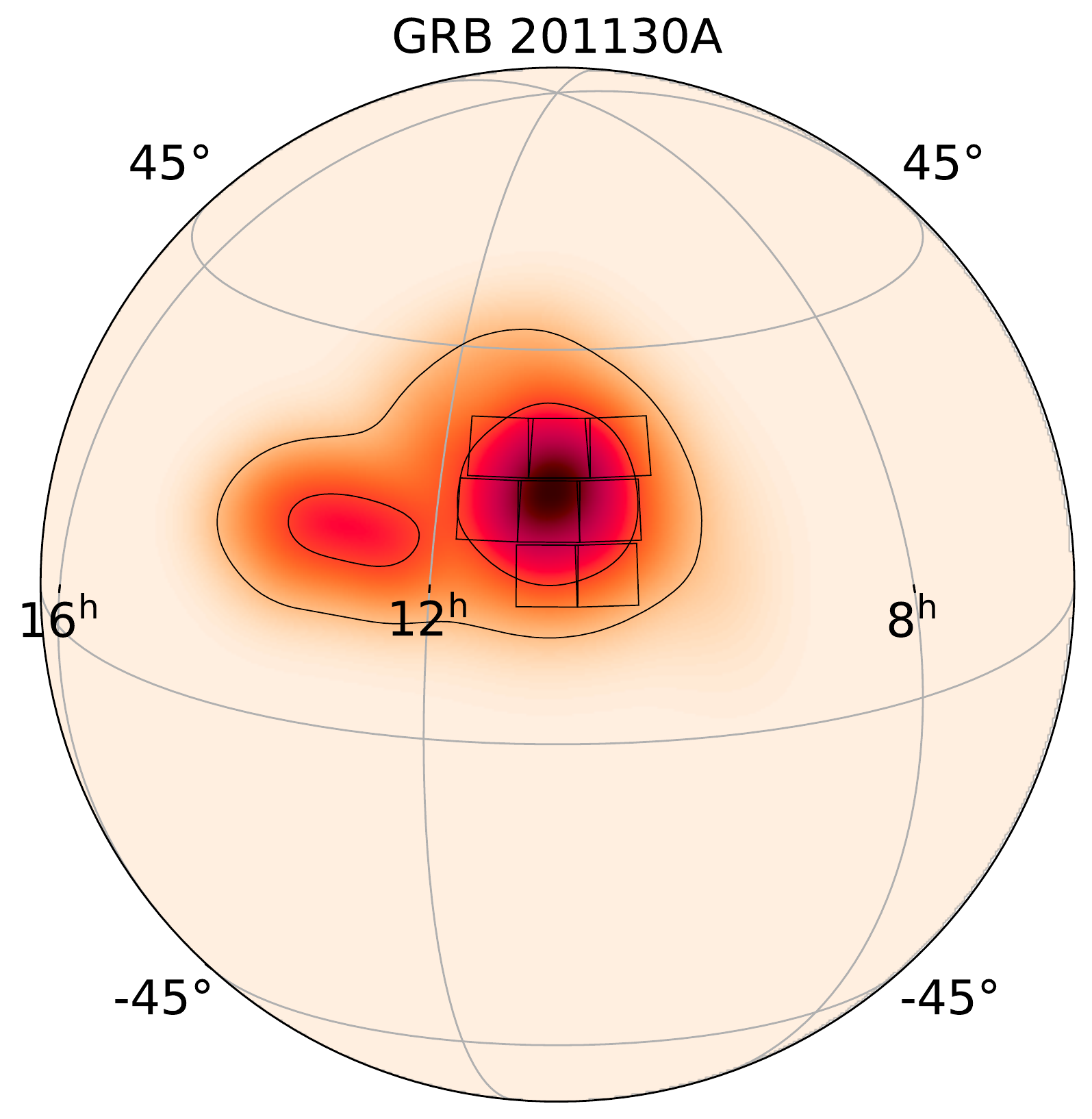}
\caption{ Coverage of four \ac{ZTF} triggers and their \textit{Fermi} \ac{GBM} localization regions. From top to bottom and left to right, the skymaps of GRB 180913, GRB 181126, GRB 200514, and GRB 201130 are shown along the $\approx$ 47 deg$^2$ \ac{ZTF} tiles (black quadrilaterals). The 50\% and 90\% credible regions are shown as black contours and the sources discovered during the \ac{ZTF} trigger as white stars (details in Section~\ref{sec:sgrb_description}). Note that for GRB 200514, we tiled the preliminary region, which was offset from the final localization. The grid shows the Right Ascension in hours and the Declination in degrees.}
 \label{fig:skymap2}
\end{figure*}

\begin{figure}[t]
\includegraphics[width=3.5in]{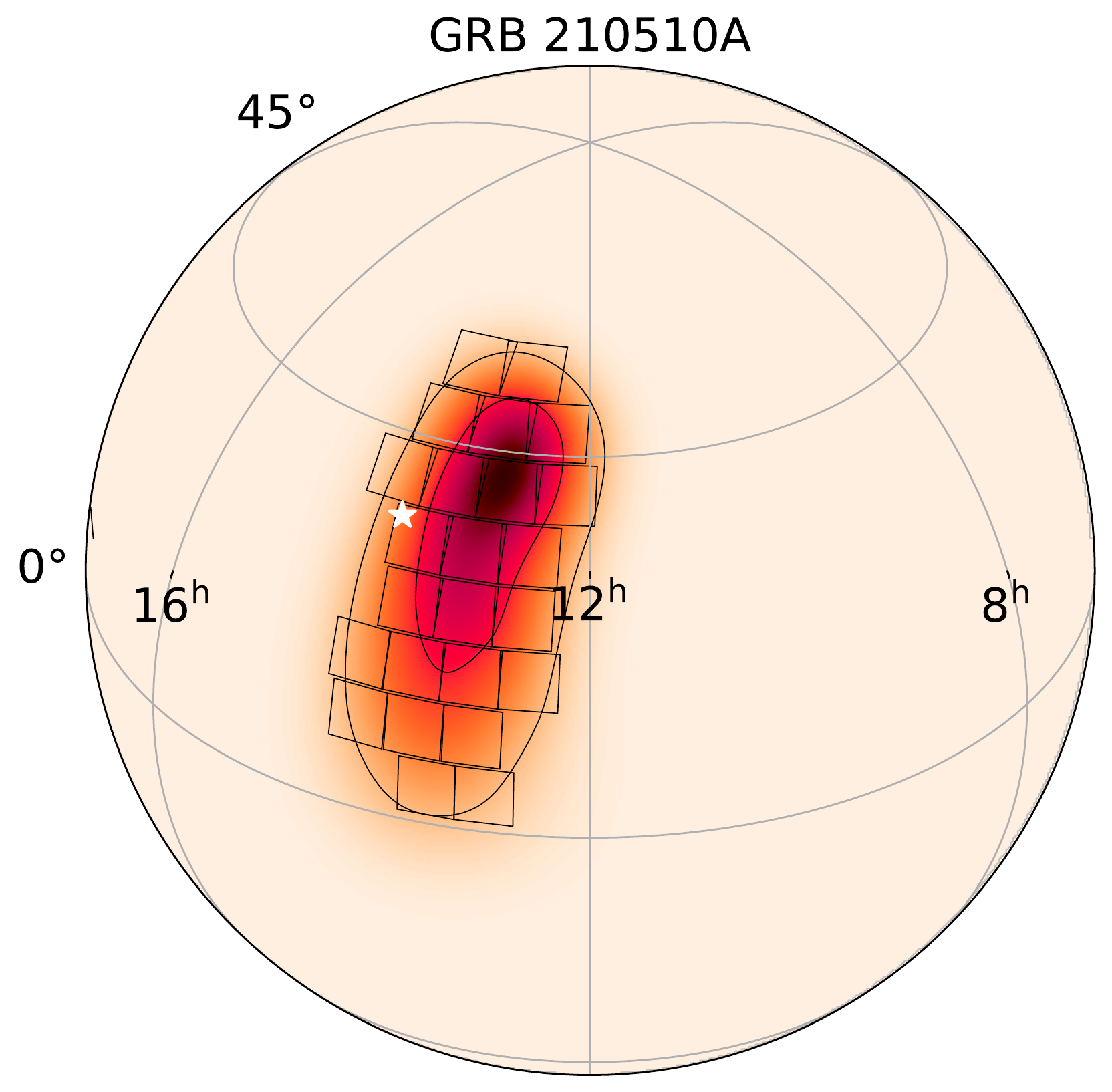}
\caption{ Coverage of the \ac{ZTF} trigger and \textit{Fermi} \ac{GBM} localization region of GRB 210510, along the $\approx$ 47 deg$^2$ \ac{ZTF} tiles (black quadrilaterals). The 50\% and 90\% credible regions are shown as black contours and the source discovered during the \ac{ZTF} trigger as white star (details in Section~\ref{sec:sgrb_description}). The grid shows the Right Ascension in hours and the Declination in degrees.}
\label{fig:skymap3}
\end{figure}

 We followed up on 10 \emph{Fermi}-\ac{GBM} \acp{SGRB}, and we show 9 skymaps and their corresponding ZTF footprints in Fig. \ref{fig:skymap1}, \ref{fig:skymap2}, and \ref{fig:skymap3}. Please refer to \citealt{ahumada2021} for details on GRB 200826A, the only short duration GRB followed up during our campaign that is not shown here. As listed in Table \ref{table:GRBs}, all of the events span more than 100\, deg$^2$, which is the average localization region covered during previous LGRBs searches \citep{SiKa2015}. Moreover, in many cases, the 90\% credible region (C.R.) spans more than 1000\, deg$^2$, which is challenging even for a 47\, deg$^2$ field of view instrument such as \ac{ZTF}. 

Triggering ToO observations for survey instruments like ZTF and Palomar Gattini-IR \citep{gattini} halts their ongoing survey observations and redirects them to observe only certain fields as directed by an observation plan. We have used \texttt{gwemopt} \citep{CoTo2018,CoAn2019}, a code intended to optimize targeted observations for gravitational wave events, to achieve an efficient schedule for our \ac{TOO} observations. The similarities between \ac{LVC} and GBM skymaps allow us to apply the same algorithm, which involves slicing the skymap into the predefined ZTF tiles and determining the optimal schedule by taking into consideration the observability windows and the need for a repeated exposure of the fields. In order to prioritize the fields with the highest enclosed probability, we used the ``greedy'' algorithm described in \cite{CoTo2018} and \cite{almualla2020scheduling}. As \texttt{gwemopt} handles both synoptic and galaxy-targeted search strategies, we employed the former to conduct observations with some of our facilities, Palomar Gattini-IR, GROWTH-India and ZTF, and the latter for scheduling observations with the Kitt Peak EMCCD Demonstrator (KPED; \citealt{Coughlin2018}).

\subsection{Optical follow-up}
Following the identification of candidate counterparts with ZTF, subsequent optical follow-up of these transients is required to characterize and classify them. For the candidates that met the requirements described in section \ref{sec:candidates}, mainly that they showed interesting light-curve history and magnitude evolution, we acquired additional data. To obtain these data, the \ac{GROWTH} multi-messenger group relies on a number of telescopes around the globe. Most of these facilities are strategically located in the Northern Hemisphere, enabling continuous follow-up of \ac{ZTF} sources. The follow-up observations included both photometric and spectroscopic observations. Even though the spectroscopic classification is preferable, photometry was essential to rule out transients, based on their color evolution and fading rates. The telescopes involved in the photometric and spectroscopic monitoring are briefly described in the following paragraphs. 

We used the \ac{KPED} on the Kitt Peak 84 inch telescope \citep{Coughlin2018} to obtain photometric data. The \ac{KPED} is an instrument mounted on a fully robotic telescope and it has been used as a single-band optical detector in the Sloan $g-$ and $r-$ bands and Johnson \textit{UVRI} filters. The \ac{FOV} is 4.4$^\prime$ $\times$ 4.4$^\prime$ and the pixel size is 0.259$^{\prime\prime}$. 

Each candidate scheduled for photometry was observed in the $g$- and $r$- band for 300\,s. The data taken with KPED are then dark subtracted and flat-field calibrated. After applying astrometric corrections, the instrumental magnitudes were determined using Source Extractor \citep{Bertin}. To calculate the apparent magnitude of the candidate, the zero-point of the field is calibrated using \ac{PS1} and \ac{SDSS} stars in the field as standards. Given the coordinates of the target, an on-the-fly query to PAN-STARRS1 and SDSS retrieves the stars within the field that have a minimum of 4 detections in each band. 

Additionally, sources were photometrically followed-up using the \ac{LCOGT} (PI: Coughlin, Andreoni) \citep{Brown_2013}. We used the 1-m and 2-m telescopes to schedule sets of 300\,s in the $g$-, $r$- and $i$-band. The \ac{LCOGT} data come already processed and in order to determine the magnitude of the transient, the same \ac{PS1}/\ac{SDSS} crossmatching strategy used for \ac{KPED} was implemented for \ac{LCOGT} images.  
 
We used the \ac{SEDM} on the Palomar 60-inch telescope \citep{BlNe2018} to acquire $g$-, $r$-, and $i$- band imaging with the Rainbow Camera on SEDM in 300\,s exposures. Images were then processed using a python-based pipeline that performs standard photometric reduction techniques and uses an adaptation of \texttt{FPipe} (Fremling Automated Pipeline; described in detail in \citealt{FrSo2016}) for difference imaging. Moreover, we employed the Integral Field Unit (IFU) on \ac{SEDM} to observe targets brighter than m$_{AB} < 19$ mag. Each observation is reduced and calibrated using the \texttt{pysedm} pipeline \citep{RiNe2019}, which applies standard calibrations using standards taken during the observing night. Once the spectra are extracted we use the \texttt{SuperNova} \texttt{IDentification}\footnote{\href{https://people.lam.fr/blondin.stephane/software/SNID/}{https://people.lam.fr/blondin.stephane/software/SNID/}} software \citep[\texttt{SNID};][]{2007ApJ...666.1024B} for spectroscopic classification.

We obtained spectra for six candidates using the \ac{DBSP} on the Palomar 200-inch telescope during  classical observing runs. The data were taken using the 1.5\arcsec slit and reduced following a custom PyRAF pipeline\footnote{https://github.com/ebellm/pyraf-dbsp}~\citep{BeSe2016}.

The other telescopes used for photometric follow-up are the \ac{GIT} in Hanle, India, the Liverpool Telescope \citep{Steele} in La Palma, Spain, and the Akeno telescope \citep{kotani2007} in Japan. The requested observations in the $g$-, $r$- and $i$-band varied between 300s\, and 600s\, depending on the telescope.

We obtained spectra with the DeVeny Spectrograph at the \ac{LDT} \citep{MacFarlane} and the 10m Keck \ac{LRIS} \citep{oke1995keck}. We reduced these spectra with PyRAF following standard long-slit reduction methods. 

We used the Gemini Multi-Object Spectrograph (GMOS-N) mounted on the Gemini-North 8-meter telescope on Mauna Kea to obtain photometric and spectroscopic data (P.I. Ahumada, GN-2021A-Q-102). Our standard photometric epochs consisted of four 180s exposures in $r$-band to measure the fading rate of the candidates, although we included $g$-band when the color was relevant. These images were processed using DRAGONS \citep{dragons} and the magnitudes were derived after calibrating against \ac{PS1}. When necessary and possible, we used \ac{PS1} references to subtract the host, using HOTPANTS. For spectroscopic data, our standard was four 650\,s exposures using the 1\arcsec long-slit and the R400 grating and we used PyRAF standard reduction techniques to reduce the data. 

\section{Candidates}
\label{sec:candidates}

After a given \ac{ZTF} observation finishes, the resulting image is subtracted to a reference image of the field \citep{MaLa2019,zogy}. The latter process involves a refined PSF adjustment and a precise image alignment in order to perform the subtraction and determine flux residuals. Any $5\sigma$ difference in brightness creates an `\textit{alert}' \citep{2019PASP..131a8001P}, a package with information describing the transient. The alerts include the magnitude of the transient, proximity to other sources and its previous history of detections among other features. 
 \ac{ZTF} generates around $10^5$ alerts per night of observation, which corresponds to $\sim 10\%$ of the estimated Vera Rubin observatory alert rate. The procedure to reduce the number of alerts from $\sim 10^{5}$ to a handful of potential optical \ac{SGRB} counterparts is described in this section. 

In general terms, the method involves a rigid online alert filtering scheme that significantly reduces the number of sources based on image quality features. Then, the selection of candidates takes into consideration the physical properties of the transient (i.e. cross-matching with AGN and solar system objects), as well as archival observations from different surveys. After visually inspecting the candidates that passed the preliminary filters, scientists in the collaboration proceed to select sources based on their light-curves, color and other features (i.e. proximity to a potential host, redshift of the host, etc.). This method allows us to recover objects that are later scheduled for further follow-up.

The candidate selection and the follow-up are coordinated via the GROWTH marshal \citep{Kasliwal2018} and lately through the open-source platform and alert broker Fritz\footnote{\url{https://github.com/fritz-marshal/fritz}}.  

\subsection{Detection and filtering}

In the searches for the optical counterpart for \acp{SGRB}, we query the ZTF data stream using the GROWTH marshal \citep{Kasliwal2018}, the \texttt{Kowalski} infrastructure \citep{kowalski}\footnote{\url{https://github.com/dmitryduev/kowalski}}, the \texttt{NuZTF} pipeline \citep{2021NatAs...5..510S,robert_stein_2021_5758176} built using \texttt{Ampel} \citep{ampel} \footnote{\url{https://github.com/AmpelProject}}, and Fritz. The filtering scheme restricted the transients to those with the following properties:

\begin{itemize}
\item  \textbf{Within the skymap:} To ensure the candidates are in the GBM skymap, we implemented a cone search in the GBM region with \texttt{Kowalski} and \texttt{Ampel}. With the GROWTH marshal approach, we retrieve only the candidates in the fields scheduled for \ac{TOO}. We note that a more refined analysis on the coordinates of the candidates is done after this automatic selection.
\item  \textbf{Positive subtraction:} After the new image is subtracted, we filter on the sources with a positive residual, thus the ones that have brightened.
\item  \textbf{It is real:} To distinguish sources that are created by ghosts or artifacts in the CCDs, we apply a random-forest model \citep{Mahabal2019} that was trained with common artifacts found in the \ac{ZTF} images. We restrict the Real-Bogus score to $> 0.25$ as it best separates the two populations. For observations that occurred after 2019, we used the improved deep learning real-bogus score \texttt{drb} and we set the threshold to sources with \texttt{drb} score $> 0.15$ \citep{kowalski}.
\item \textbf{No point source underneath:} To rule out stellar variability we require the transient to have a separation of 3\arcsec\, from any point source in the \ac{PS1} catalog based on \cite{TaMi2018}. 
\item  \textbf{Two detections:} We require a minimum of two detections separated by at least 30\,min. This allows us to reject cosmic rays and moving solar system objects. 
\item  \textbf{Far from a bright star:} To further avoid ghosts and artifacts, we require the transient to be $>$ 20\arcsec\, from any bright ($m_\textrm{AB} < 15$ mag) star. 
\item  \textbf{No previous history:} As we do not expect the optical counterpart of a \ac{SGRB} to be a periodic variable source, we restrict our selection to only sources that are detected after the event time and have no alerts generated for dates prior to the GRB.  
\end{itemize}

As a reference, this first filtering step reduced the total number of sources to a median of $\sim 0.03\%$ of the original number of alerts. The breakdown of each filter step is shown in Table \ref{table:filtering}. A summary of the numbers of followed-up objects for each trigger is in Table \ref{table:GCNs} and the details of the filtering scheme are described below. More than 3 $\times 10^{5}$ alerts were generated during the 9 \ac{TOO} triggers, while $\sim$80 objects were circulated in the \ac{GCN}.

\begin{deluxetable*}{ccccccccc}
\tablecaption{\label{table:filtering} Summary of the efficiency of our vetting strategy. For each GRB we list the number of alerts that survives after a given filtering step. The first column (SNR$>$5) shows the total number of alerts in the GRB map. The next column displays the number of alerts that show an increase in flux (Positive subtraction). The 'Real` column shows the number of sources considered as real using either the real-bogus index(RB) or \texttt{drb} scores. We set the thresholds to RB$>$0.25 and \texttt{drb}$>$0.5. The next columns show the number of sources that are not related to a point source, nor close to a bright star, to avoid artifacts. To avoid moving objects, we show the number of sources with two detections separated by at least 30 min. The last column shows the number of sources we circulated as potential candidates for each trigger. For each step, we calculate the median reduction of alerts and list this number at the end of each column.}
 \tablehead{\colhead{GRB} & \colhead{SNR$>$5} & \colhead{\begin{tabular}{@{}c@{}}Positive \\ subtraction\end{tabular}} &  \colhead{Real} & \colhead{\begin{tabular}{@{}c@{}}Not star \\ underneath\end{tabular}} & \colhead{\begin{tabular}{@{}c@{}}Far from  \\ bright star\end{tabular}} & \colhead{\begin{tabular}{@{}c@{}}Two \\ detections\end{tabular}} &\colhead{\begin{tabular}{@{}c@{}}Circulated \\ in GCNs\end{tabular}}  } 
 \startdata
GRB 180523B & 67614 & 17374 & 12117 & 687   & 669   & 297 &14 \\ 
GRB 180626C & 10602 & 5040  & 4967  & 1582  & 1377  & 214 &1 \\ 
GRB 180715B & 33064 & 7611  & 7515  & 6941  & 5509  & 104 &14 \\
GRB 180728B & 18488 & 1450  & 1428  & 859   & 739   & 51  &7 \\ 
GRB 180913A & 25913 & 12105 & 12077 & 6284  & 5145  & 372 &12 \\
GRB 181126B & 40342 & 30455 & 30416 & 22759 & 21769 & 340 &11 \\ 
GRB 200514B & 20610& 10983& 10602& 4502& 4422& 1346 &14 \\ 
GRB 200826A & 13488 & 8142 & 7744 & 3892 & 3785 & 464 & 14\\ 
GRB 201130A & 1972& 1045& 990& 647& 637& 43  & 0\\ 
GRB 210510A & 41683 & 27229& 28940 & 16977 & 16973 & 1562   & 1 \\
\hline
Median reduction && 50.27\% &  48.53\% &  23.05 \% &  20.66\% &   1.73\% &   0.03\% \\
\enddata
\end{deluxetable*}

\subsection{Scanning and selection}

Generally, after the first filter step, the number of transients is reduced to a manageable amount $\sim O(100)$. These candidates are then cross-matched with public all-sky surveys such as {\it \aclu{WISE}}  \citep[\textit{WISE};][]{2013wise.rept....1C}, \aclu{PS1} \citep[PS1; ][]{2016arXiv161205560C}, \aclu{SDSS} \citep[SDSS; ][]{dr16sdss}, the \aclu{CRTS} \citep[CRTS; ][]{catalina}, and the \aclu{ATLAS} \citep[ATLAS; ][]{Ton2011}. We use the {\it \ac{WISE}} colors to rule out candidates, as \acp{AGN} are located in a particular region in the {\it \ac{WISE}} color space \citep{WrEi2010,StAs2012}. If a candidate has a previous detection in \ac{ATLAS} or has been reported to the \ac{TNS} before the event time it is also removed from the candidate list. We additionally crossmatch the position of the candidates with the \ac{MPC} to rule out any other slow moving object. We use the \ac{PS1} DR2 \footnote{\url{https://catalogs.mast.stsci.edu/panstarrs/}} to query single detections at the location of the transients, and we use this information to rule out sources based on serendipitous previous activity. 

One of the most important steps in our selection of transients is the rejection of sources using \ac{FP} on ZTF images. For this purpose we run two \ac{FP} pipelines: ForcePhotZTF\footnote{\url{https://github.com/yaoyuhan/ForcePhotZTF}} \citep{yao2019ztf} and the ZTF \ac{FP} pipeline \citep{MaLa2019}. We limit our search to 100 days before the burst and reject sources with consistent $\geq$ 4$\sigma$ detections.

\begin{figure*}[!htb]
\begin{minipage}[b]{0.49\linewidth}
\centering
    \includegraphics[width=\linewidth]{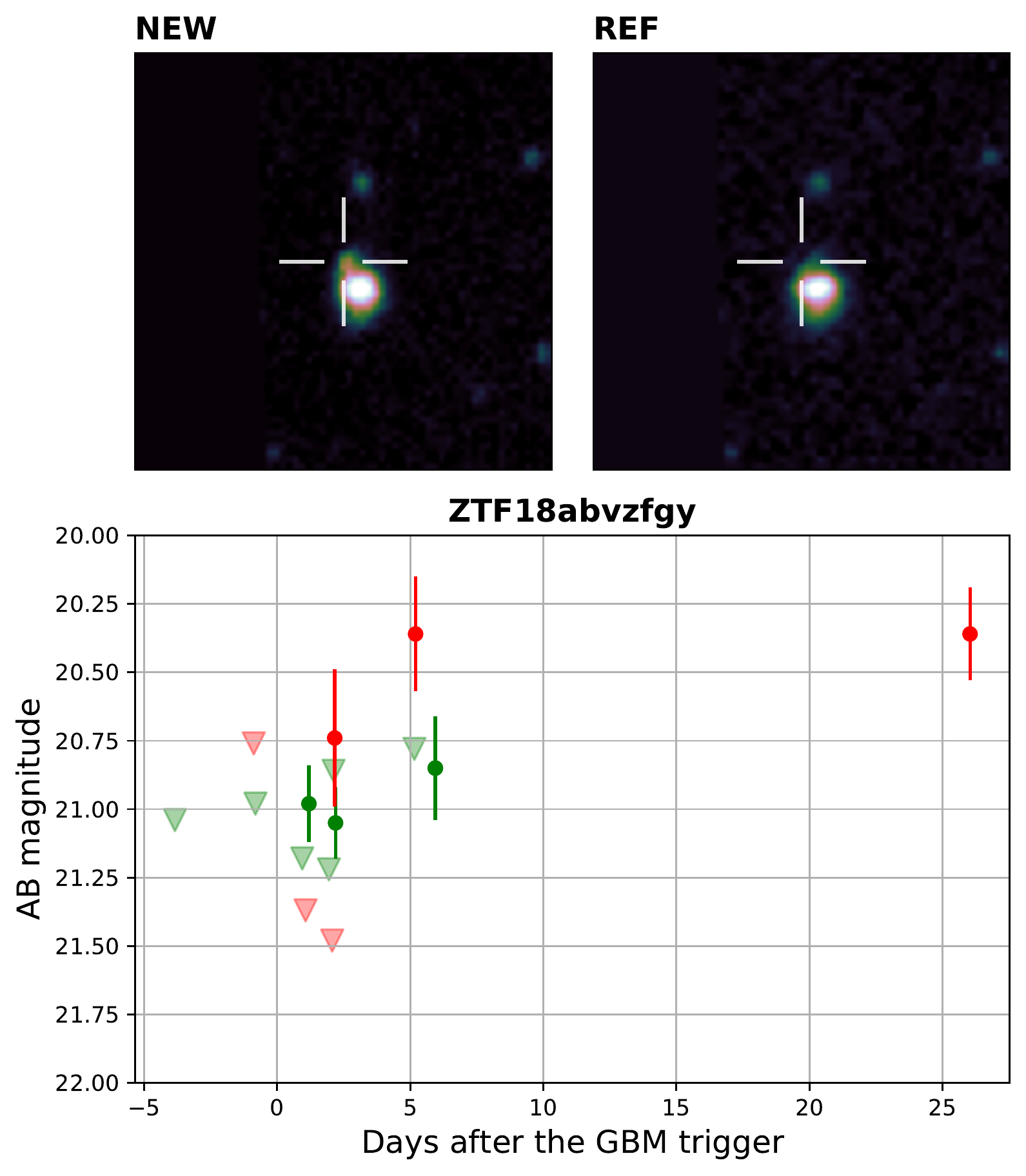}
\end{minipage}%%
\begin{minipage}[b]{0.49\linewidth}
\centering
    \includegraphics[width=\linewidth]{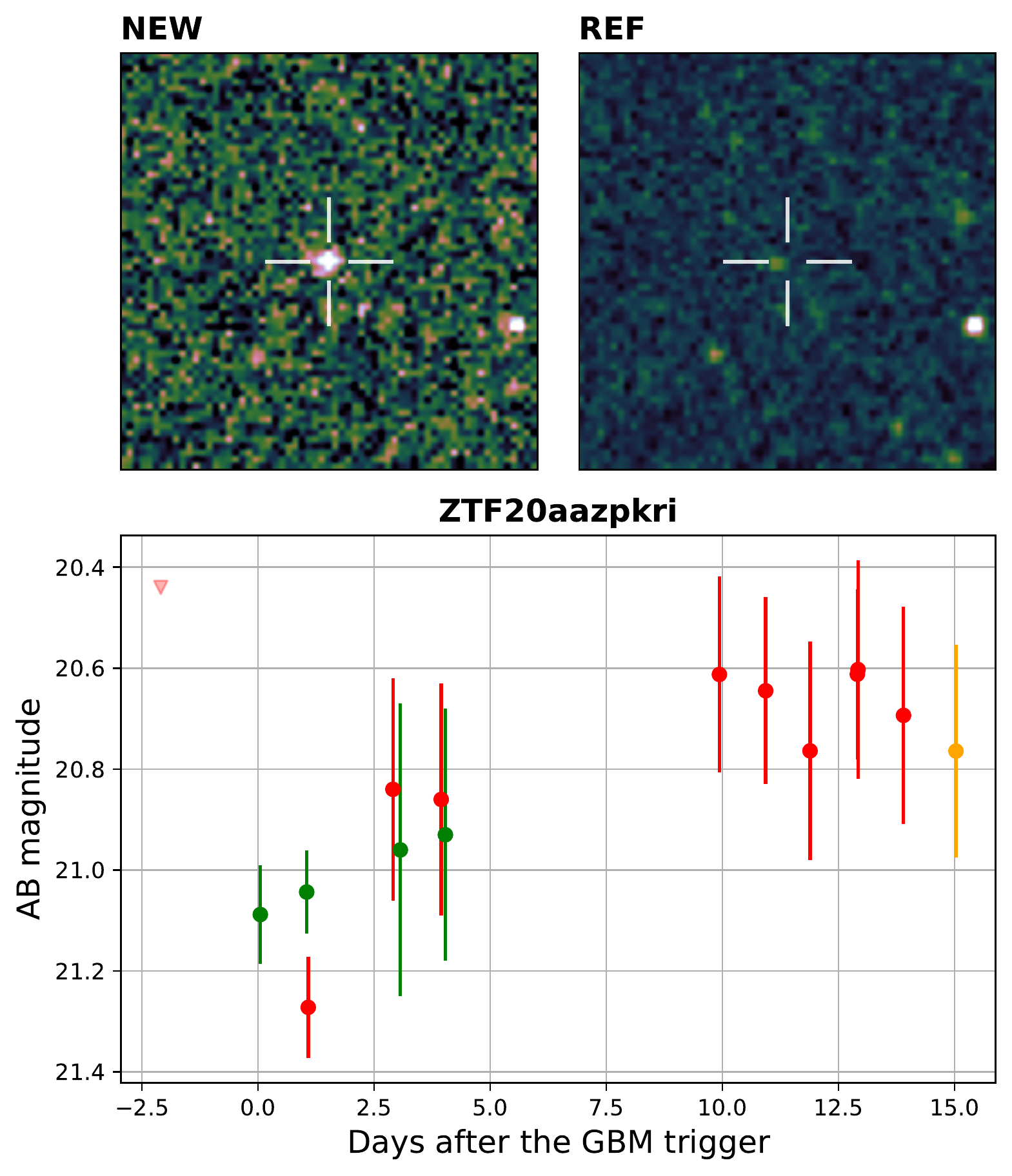}
\end{minipage} 
\caption{Examples of light-curves and cut-outs for candidates that passed our filtering criteria. Candidate ZTF18abvzfgy (candidate counterpart to GRB 180913A) in the left panel and ZTF20aazpkri (candidate counterpart to GRB 200514B) in the right panel. The observations in $g-$ and $r-$ band are plotted in green and red colors respectively. Filled circles represent \ac{ZTF} detections, while the 5$\sigma$ upper limits are shown as triangles in the light-curve. The top half of each panel shows the discovery image on the left and the reference image on the right. In the 0.7 sq. arcmin cutouts, north is up and east is to the left. A cross marks the location of the transient. }\label{fig:lc_transients}
\end{figure*}

Finally, we manually scan and vet candidates passing those cuts, referring to cutouts of the science images, photometric decay rates, and color evolution information in order to select the most promising candidates (see Fig.~\ref{fig:lc_transients}).

Detailed tables with the candidates discovered by \ac{ZTF} for the SGRB campaign are shown in Table \ref{table:candidates_all}

\subsection{Rejection Criteria}
In order to find an optical counterpart, further monitoring of the discovered transients is needed. We have taken spectra for the most promising candidates to classify them. Most of the spectra acquired correspond to bright \acp{SN} (as in Fig. \ref{fig:SNspectra}) and a few \acp{CV} and an \ac{AGN}. After the 9 \acp{SGRB} follow-ups, we obtained 19 spectra, however none of them exhibited \ac{KN} features. We have used the `Deep Learning for the Automated Spectral Classification of Supernovae and Their Hosts' or \texttt{dash} \citep{dash} to determine the classification of the candidates with \ac{SN} spectral features. \acp{CV} were recognized as they show H features at redshift $z$ = 0.

For the sources that do not have spectra available, we monitored their photometric evolution with the facilities described in Section \ref{sec:ObsNData}. Even though the photometric classification cannot be entirely conclusive, there are characteristic features shared between afterglows and KNe. On one side, afterglows are known to follow a power-law decay of the form $F \sim t^{-\alpha}$. On the other hand, most KN models \citep{Bulla2019} show evolution faster than 0.3 mag per day (\citealt{anand2020nsbh,andreoni2020constraining}). As a reference, \ot\ faded over $\sim$ 1\, mag over the course of 3\, days and other \ac{SGRB} optical counterparts have shown a rapid magnitude evolution as well \citep{Fong15,rastinejad2021}. The astrophysical events that most contaminated our sample are \acp{SN}, but they normally show a monotonic increase in their brightness during their first tens of days, to later decline at a slower rate than expected for afterglows or KNe. Other objects like slow-moving asteroids and flares are less common and can be removed inspecting the images or performing a detailed archival search in ZTF and other surveys. 

To illustrate the photometric rejection, we show two transients in Fig. \ref{fig:lc_transients} with no previous activity in the ZTF archives previous to the \ac{SGRB}. As their magnitude evolution in both $r-$ and $g-$ band does not pass our threshold, we conclude that they are not related to the event. This process was repeated for all candidates without spectral information, using all the available photometric data from ZTF and partner telescopes.

\section{\ac{SGRB} events} \label{sec:sgrb_description}

\subsection{GRB 180523B}
The first set of \ac{TOO} observations of this program was taken 9.1 hours after GRB 180523B (trigger 548793993). We covered $\sim$ 2900\, deg$^2$, which corresponds to 60\% of the localization region after accounting for chip gaps in the instrument \citep{gcn22739}. 
The median 5$\sigma$ upper limit for an isolated point source in our images was r $>$ 20.3\,mag and g $>$ 20.6\, mag and after 2\, days of observations we arrived at 14 viable candidates that required follow-up.  We were able to spectroscopically classify 4 transients as \acp{SN} and photometrically follow-up sources with \ac{KPED} to determine that the magnitude evolution was slower than our threshold. This effort was summarized in \cite{CoAh19sgrb..131d8001C} and the list of transients discovered is displayed in Table \ref{table:candidates_all}.

\subsection{GRB 180626C}
The \ac{SGRB} GRB 180626C (\emph{Fermi} trigger 551697835) came in the middle of the night at Palomar. We started observing after 1.5\, hours and were able to cover 275 deg$^2$ of the \ac{GBM} region. The localization, and hence the observing plan, was later updated as the region of interest was now the overlap between the \emph{Fermi} and the newly arrived \ac{IPN}\footnote{\url{http://www.ssl.berkeley.edu/ipn3/index.html}} map. The observations covered finally 230 deg$^2$, corresponding to  87\% of the intersecting region. After two nights of observations, with a median 5-sigma upper limit of r $>$ 21.1 mag and g $>$ 21.0 mag, only one candidate was found to have no previous history of evolution and be spatially coincident with the \ac{SGRB} \citep{gcn22871}.

The transient ZTF18aauebur was a rapidly evolving transient that faded from
g = 18.4 to g = 20.5 in 1.92 days. This rapid evolution continued during the following months, fluctuating between r $\sim$ 18 mag and r $\sim$ 19 mag.  It was interpreted as a stellar flare, as it is located close to the Galactic plane and there is an underlying source in the \ac{PS1} and {\it \ac{GALEX}} \citep{galex} archive. Additionally, its \ac{SEDM} spectrum showed a featureless blue spectrum and H$\alpha$ absorption features at redshift z = 0, so it is an unrelated Galactic source. The rest of the candidates can be found in Table \ref{table:candidates_all}.

\subsection{GRB 180715B}
We triggered \ac{TOO} observations to follow-up GRB 180715B (trigger 553369644) 10.3\, hours after the \ac{GBM} detection. We managed to observe $\sim$ 36\% of the localization region which translates into 254 deg$^2$. The median limiting magnitude for these observations was r $>$ 21.4 mag and g $>$ 21.3 mag. 

During this campaign, we discovered 14 new transients \citep{gcn22969} in the region of interest. We were able to spectroscopically classify 2 candidates using instruments at the \ac{P60} and \ac{P200}. The \ac{SEDM} spectrum of ZTF18aauhpyb showed a stellar source with Balmer features at redshift z = 0 and a blue continuum. The \ac{DBSP} spectrum of ZTF18abhbfqf was best fitted by a \ac{SN} Ia-91T. We show the rejection criteria used to rule-out associations with the \ac{SGRB} in Table \ref{table:candidates_all}. Generally, most candidates showed a slow magnitude evolution. Furthermore, three candidates (ZTF18abhhjyd, ZTF18abhbfoi and ZTF18abhawjn) matched with an \ac{AGN} in the Milliquas \citep{milliquas} catalog. A summary of the candidates can be found in Table \ref{table:candidates_all}.

\subsection{GRB 180728B}
The \ac{TOO} observations of GRB 180728B (trigger
554505003) started $\sim$ 8\, hours after the \emph{Fermi} alert, however, it did not cover the later updated \ac{IPN} localization. The following night and 31\, hours after the \emph{Fermi} detection we managed to observe the joint \ac{GBM} and \ac{IPN} localization, covering 334 deg$^2$ which is $\sim$ 76\% of the error region. The median upper limits for the scheduled observations were r $>$ 18.7 mag and g $>$ 20.0 mag \citep{gcn23379}.  As a result of these observations, no new transients were found.

\subsection{GRB 180913A}
We triggered \ac{TOO} observations with \ac{ZTF} to follow-up the \emph{Fermi} event GRB 180913A (trigger 558557292) about $\sim$ 8\, hours after the GBM detection. The first night of observations covered 546\, deg$^2$. The schedule was adjusted as the localization improved once the \ac{IPN} map was available. During the second night we covered 53\% of the localization, translated into 403\, deg$^2$. After a third night of observations, 12 transients were discovered and circulated in \cite{gcn23324}. The median upper limits for this set of observations were r $>$ 21.9 and g $>$ 22.1 mag.

We obtained a spectrum of ZTF18abvzfgy with \ac{LDT}, a fast rising transient ($\Delta m / \Delta t \sim -0.2$ mag per day) in the outskirts of a potential host galaxy. It was classified as a \ac{SN} Ic at a redshift of z = 0.04. The rest of the transients were follow-up photometrically with KPED and LCO, but generally showed a flat evolution. The candidate ZTF18abvzsld had previous \ac{PS1} detections, thus ruling it out as a SGRB counterpart. The rest of the candidates are listed in Table \ref{table:candidates_all}. 

\subsection{GRB 181126B}
The last \ac{SGRB} we followed-up before the start of the 2019 O3 LIGO/Virgo observing run was of the \emph{Fermi}-\ac{GBM} event GRB 181126B (trigger 564897175). As this event came during the night at the \ac{ZTF} site, the observations started $\sim$ 1.3\, hours after the \emph{Fermi} alert, and we were able to cover 1400\, deg$^2$, close to 66\% of the \ac{GBM} localization. After the \ac{IPN} localization was available the next day, the observations were adjusted and we used \ac{ZTF} to cover 709 deg$^2$, or $\sim$ 76\% of the overlapped region. The mean limiting magnitude of the observations was r $>$ 20.8 mag \citep{gcn23515}. After processing the data, we discovered 11 new optical transients timely and spatially coincident with the \ac{SGRB} event. We took spectra of 7 of them with the Keck \ac{LRIS}, discovering 6 \acp{SN} (ZTF18acrkkpc, ZTF18aadwfrc, ZTF18acrfond, ZTF18acrfymv, ZTF18acptgzz, ZTF18acrewzd) and 1 stellar flare (ZTF18acrkcxa). All of the candidates  are listed in Table \ref{table:candidates_all}, and none of them showed rapid evolution.  

\subsection{GRB 200514B}
We resumed the search for SGRB counterparts with ZTF once LIGO/Virgo finished O3. On 2020-05-14 we used ZTF to cover over 519.3 deg$^2$ of the error region of GRB 200514B (trigger 611140062). This corresponds to $\sim 50\%$ of the error region. After the first night of observations, 7 candidates passed our filters and were later circulated in \cite{gcn27737}. The observations during the following night resulted in 7 additional candidates \citep{gcn27745}. The depth of these observations reached 22.4 and 22.2 mag in the $g-$ and $r-$band respectively. After IPN released their analysis \citep{GCN27755}, 9 of our candidates remained in the localization region. Our follow-up with ZTF and LCO showed that none of these transients evolved as fast as expected for a GRB afterglow (see Table \ref{table:candidates_all}). 

\subsection{GRB 200826A}
This burst is discussed extensively in \cite{ahumada2021}, as well as in other works \citep{Zhang2021,rossi2021,rhodes2021}. It was the only short duration GRB in our campaign with an optical counterpart association. However, despite its short duration ($T_{90}$ = 1.13s), it showed a photometric bump in the $i$-band that could only be explained by an underlying SN \citep{GCN29029,2020GCN.29029....1A}. This makes GRB 200826A the shortest-duration \ac{LGRB}  \cite{ahumada2021}.  

\subsection{GRB 201130A}  
The ZTF trigger on GRB 201130A reached a depth of r = 20.5\,mag in the first night of observations after covering 75\% of the credible region. No optical transient passed all our filtering criteria \citep{gcn28981}. 

\subsection{GRB 210510A} 
\begin{figure*}
\begin{center}
    \includegraphics[width=0.95\textwidth]{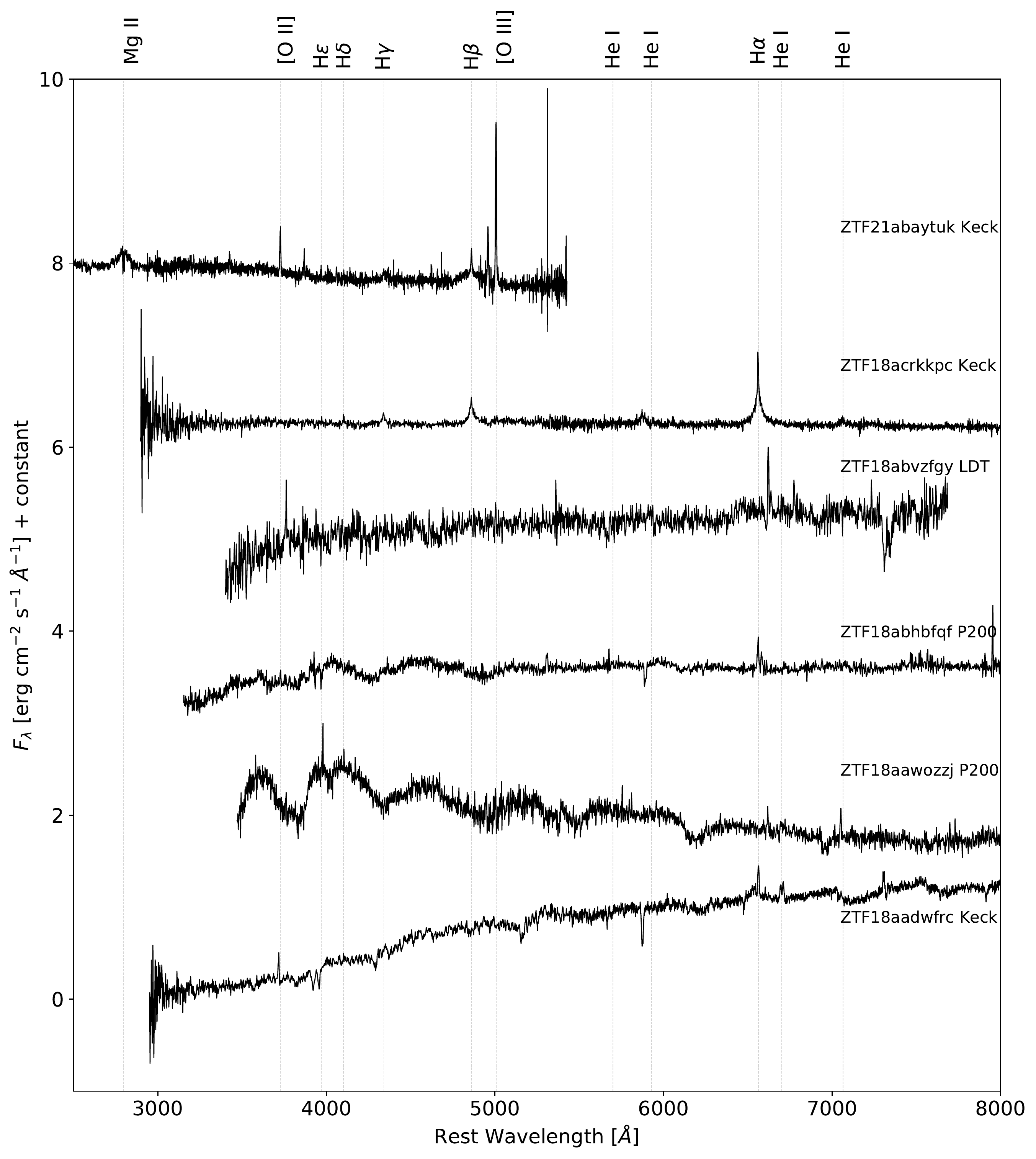}
\end{center}
 \caption{\label{fig:SNspectra} The spectra of some representative candidates. The spectrum of transient ZTF18aadwfrc was taken with the LRIS at the Keck Observatory and was classified as a SN Ia at z = 0.04. Similarly, the spectrum of ZTF18acrkkpc and ZTF21abaytuk come from Keck as well, and were classified as a SN II at z = 0.061 and as an AGN at z = 0.89 respectively. We used the DBSP at P200 to aqcuire spectra of ZTF18aawozzj and ZTF18abhbfqf, two SN Ia at redshift z = 0.095 and z= 0.11 respectively. Lastly, the spectrum of ZTF18abvzfgy was obtained with the DeVeny Spectrograph at the LDT, and using \texttt{dash}, we classified it as a SN Ic at z = 0.04. For reference, we show the Hydrogen, Helium, Magnesium, and some Oxygen lines as vertical lines.}
\end{figure*}

We triggered optical observations on GRB 210510A (trigger 642367205) roughly 10 hrs after the burst. The second night of observations helped with vetting candidates based on their photometric evolution, at least a 0.3 mag per day decay rate is expected for afterglows and KNe. The only candidate that passed our filtering criteria was ZTF21abaytuk \citep{gcn30005}, however its Keck LRIS spectrum showed H$\beta$, [O II], and [O III] emission features and Mg II absorption lines at redshift of z = 0.89 (see Table \ref{table:candidates_all} and Fig. \ref{fig:SNspectra}). Its spectrum, summed with its {\it \ac{WISE}} colors, are consistent with an AGN origin. 

\section{\ac{ZTF} upper limits}\label{sec:upperlimits}

\begin{figure*}[!htb]
  \begin{minipage}[b]{0.5\linewidth}
    \centering
    \includegraphics[width=\linewidth]{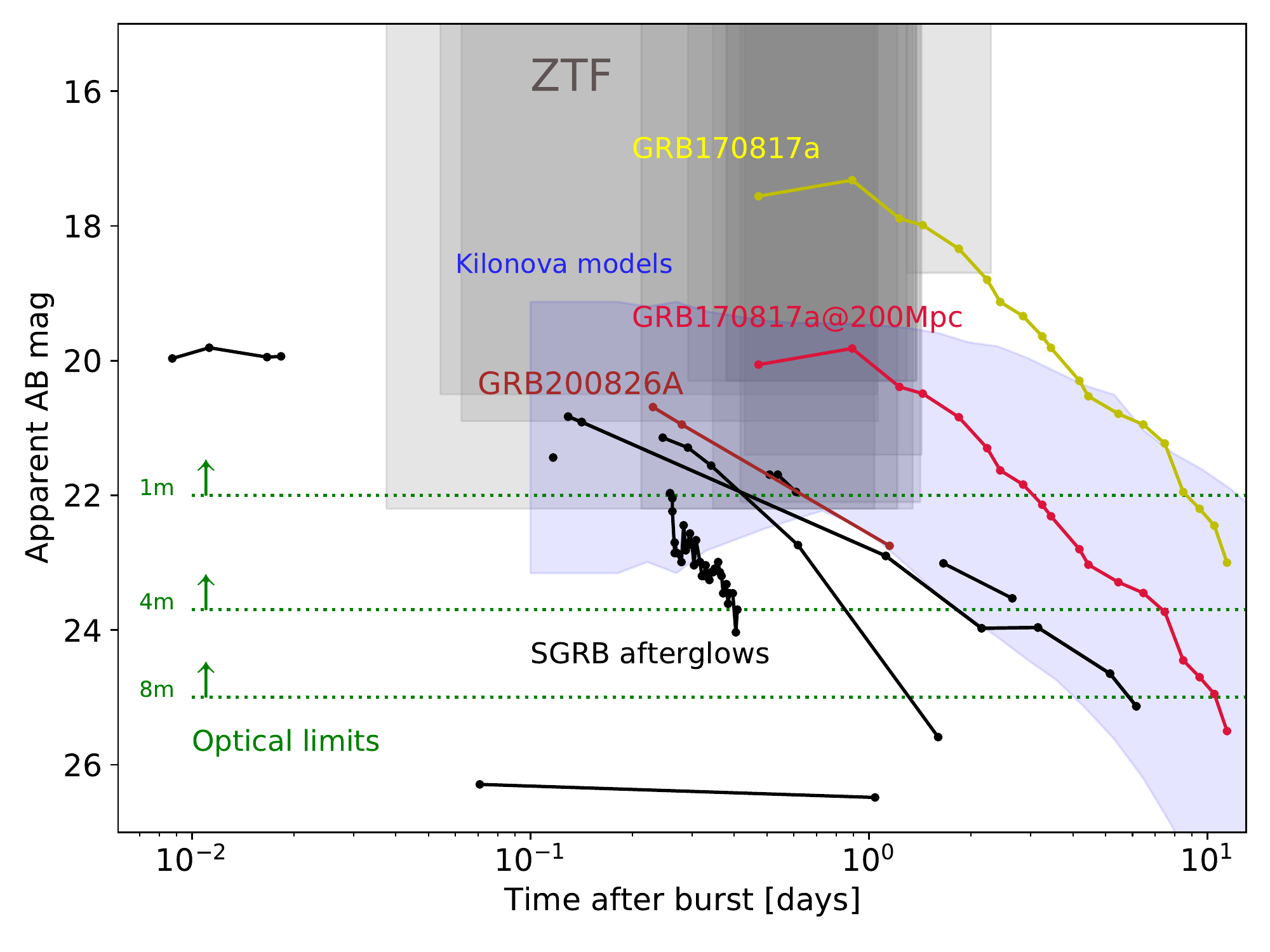}
  \end{minipage}%%
  \begin{minipage}[b]{0.5\linewidth}
    \centering
    \includegraphics[width=\linewidth]{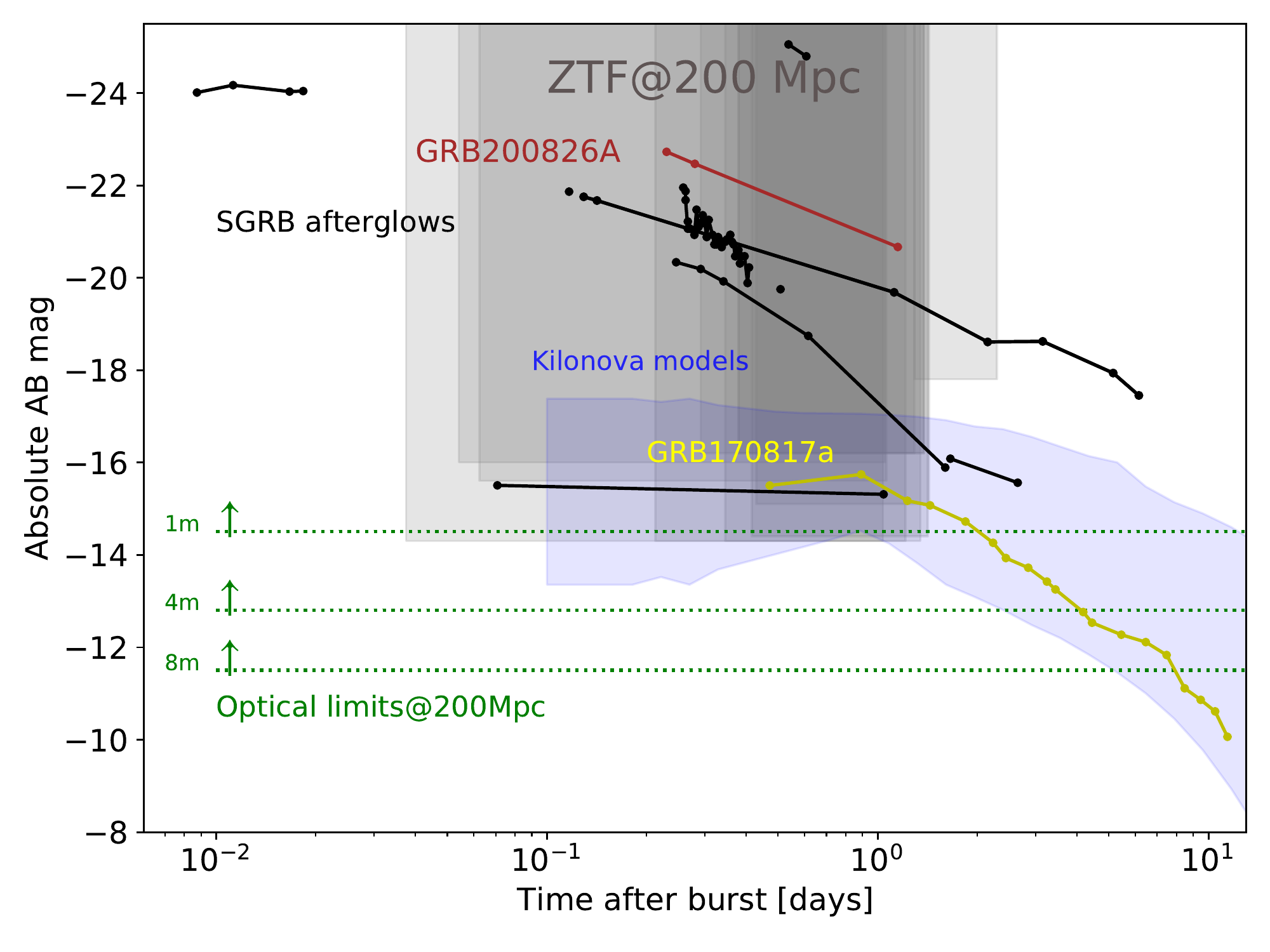}
  \end{minipage} 
\caption{\emph{(left)} The light-curves (black) of the optical counterparts of \acp{SGRB} with known redshift listed in \cite{Fong15}. The yellow light-curve is the \ot\ light-curve and the red line is the \ot\ light-curve scaled to a distance of 200\, Mpc. Each of the \ac{ZTF} search windows occupies a grey region, limited by the median limiting magnitude and the time window in which the search took place. The brown light-curve is the afterglow of GRB 200826A \citep{ahumada2021} and the blue shaded region represents the region that the KN models \citep{Bulla2019,Dietrich2020bns} occupy when scaled to 200 Mpc. The green-dotted lines represent the typical optical limits of imagers mounted at different telescopes, while the size of the telescope is annotated as a label in the plot.
\emph{(right)} The absolute magnitude of the same data plotted in the left panel. We compare their absolute magnitudes to the \ac{ZTF} magnitude limits, scaled to a fiducial distance of 200\, Mpc. Similarly, the green-dotted lines show the optical limits of different facilities, ranging in size, at 200 Mpc.}\label{fig:non-det}
\end{figure*}

It is possible to compare the search sensitivity, both in terms of depth and timescale, to the expected afterglow and kilonova light-curves. In the left panel of Fig. \ref{fig:non-det}, the median limits for \ac{ZTF} observations are shown with respect to known \textit{Swift} \ac{SGRB} afterglows with measured redshift from \cite{Fong15}. The yellow light-curve corresponds to \ot\ \citep{AbEA2017} and the red line is the same \ot\, light-curve scaled to a distance of 200\, Mpc (see below). Along with \ot, we show a collection of KN light-curves from a BNS grid \citep{Bulla2019,Dietrich2020bns} scaled to 200 Mpc. The regions of the light-curve space explored by each \ac{ZTF} trigger are represented as grey rectangles and the more opaque region corresponds to their intersection. Even though \ac{ZTF} has the ability to detect a \ot-like event and most of the KN lightcurves, most of the \ac{SGRB} afterglows observed in the past are below the median sensitivity of the telescope. On the other hand, the counterpart of the GRB 200826A would have been detected in six of our searches, even though it is on the less energetic part of the LGRB distribution. When scaled to 200 Mpc, the \ot\, light-curve overlaps with the region of five of our searches, suggesting that the combination of depth and rapid coverage of the regions could allow us to detect an \ot-like event. The searches that do not overlap with the scaled \ot\, have either fainter median magnitude upper limits ($< 20$ mag) or late starting times ($>1$ day). 

\begin{figure}[!htb]
    \centering
    \includegraphics[width=\linewidth]{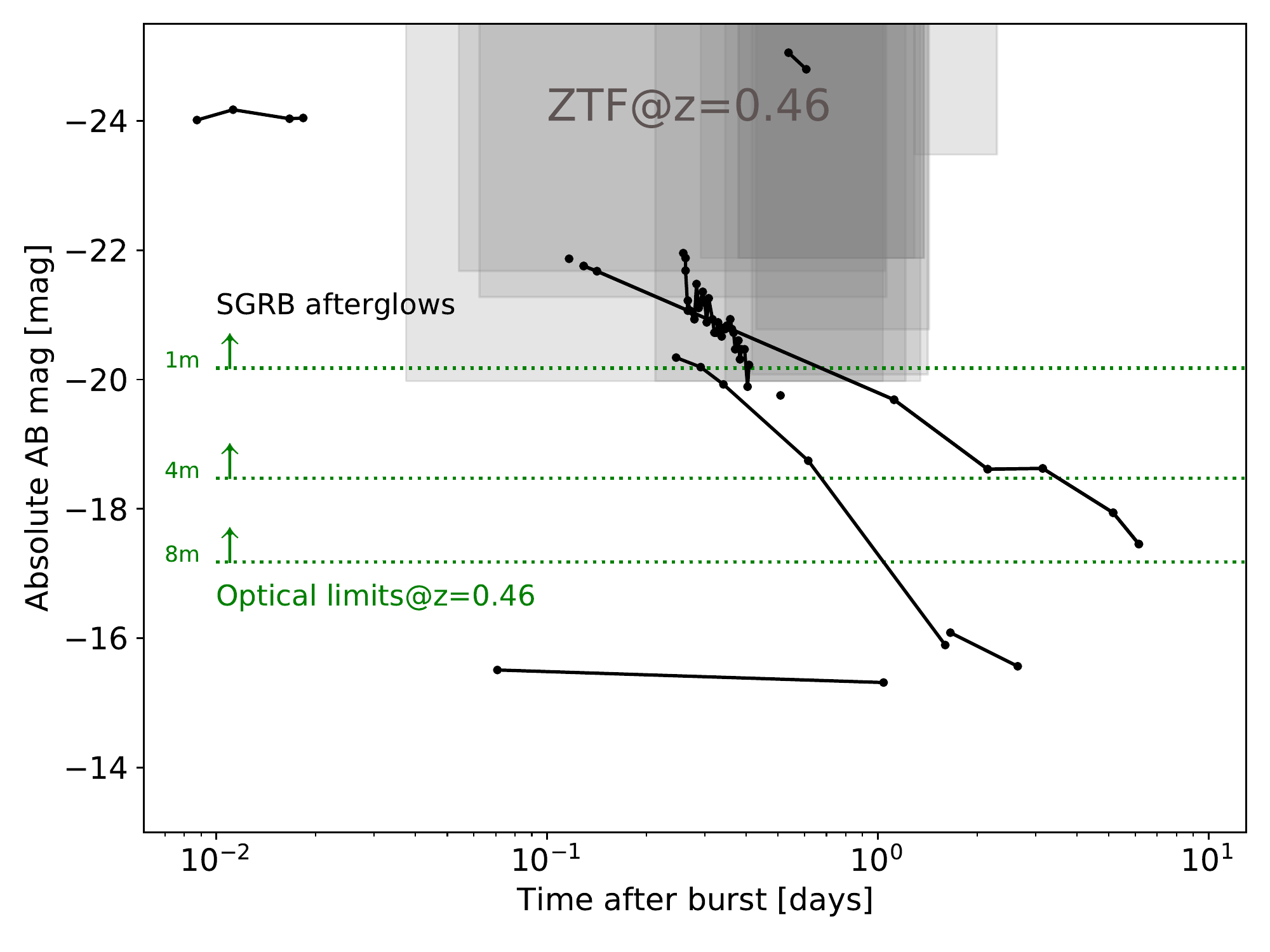}
\caption{ The absolute magnitude (black) of the optical counterparts of \acp{SGRB} with known redshift listed in \cite{Fong15}. Each of the \ac{ZTF} search windows occupies a grey region, limited by the median limiting magnitude and the time window in which the search took place. The median limiting magnitudes are scaled to the median SGRB redshift of z=0.47. The green-dotted lines represent the typical optical limits of imagers mounted at different telescopes, while the size of the telescope is annotated as a label in the plot. These limits are also scaled to the median SGRB redshift of at z = 0.47.}\label{fig:non-det-sgrb}
\end{figure}

We used the redshifts of the \acp{SGRB} optical counterparts to determine their absolute magnitudes, which is plotted in the right panel in Fig. \ref{fig:non-det}, along with GRB 200826A and \ot. In order to compare with the \ac{ZTF} searches and constrain the observations, the median \ac{ZTF} limits were scaled to a fiducial distance of 200\,Mpc, the O3 LIGO/Virgo detection horizon \citep{AbEA2018} for \ac{BNS} mergers. The range of 200\,Mpc is coincidentally approximately the furthest distance as to which \ac{ZTF} can detect a \ot-like event based on the median limiting magnitudes of this experiment. Moreover, the ZTF region covers most of the KNe models (blue shaded region) scaled to 200 Mpc. In contrast to the left panel in Fig. \ref{fig:non-det}, most of the \ac{SGRB} optical afterglows fall in the region explored by ZTF. Therefore, if any similar events happened within 200 Mpc, the current \ac{ZTF} \ac{TOO} depth plus a rapid trigger of the observations should suffice to ensure coverage in the light-curve space. Previous studies \citep{dichiara2020short} have come to the conclusion that the low rate of local SGRB is responsible for the lack of detection \ot-like transients. In fact, the probability that one of the SGRBs in our sample is within 200\,Mpc is 0.3, given the rate derived in \citet{dichiara2020short} of 1.3 SGRB within 200\,Mpc per year, assuming an average of 40 SGRBs per year. In Fig.~\ref{fig:non-det-sgrb} we show the same SGRB absolute magnitude light-curves, but in this case we compared them to the ZTF limits scaled to the median redshift of z = 0.47 from \citet{Fong15}. The ZTF search is still sensitive to SGRB afterglows at these distances within the first day after the GRB event. 

\section{Efficiency and joint probability of non-detection} \label{sec:non-det} 

In this section we determine the empirical detection efficiency for each of our searches, and use these efficiencies to calculate the likelihood of detecting a SGRB afterglow in our \ac{TOO} campaign. With this approach we are able to set limits on the ZTF's ability of detecting SGRB afterglows as a function of the redshift of the SGRB. To accomplish this, we take each GRB we followed-up and inject afterglow light-curves in the GRB maps at different redshifts. We derive efficiencies using the ZTF observing logs, since these logs contain the coordinates of each successful ZTF pointing and the limiting magnitude of each exposure. This already takes into consideration weather and other technical problems with the survey. In this section we describe the computational tools used in this endeavor and the results derived from these simulations. 

We use \texttt{simsurvey} \citep{feindt2019simsurvey} to inject afterglow-like light-curves into the GBM skymaps. We distributed the afterglows according to the GBM probability maps and within the 90\% credible region of each skymap. We slice the volume into seven equal redshift bins, from z = 0.01 to z = 2.1, and injected 7000 sources in each slice. For each injected transient, \texttt{simsurvey} employs light-curve models to derive the magnitude of the source at different times (see below for the models used). \texttt{simsurvey} uses the ZTF logs to determine if the simulated source was in an observed ZTF field and whether the transient would have been detected given the upper limits of that ZTF field.

One of the driving features of an afterglow model is its isotropic-equivalent energy, $E_{iso}$, as it sets the luminosity of the burst and hence its magnitude and light-curve. The information provided by the \textit{Fermi}-\ac{GBM} gamma-ray detections does not give insights on the distance to the event or the energies associated with the \acp{SGRB}. For this reason, and to get a sense on the $E_{iso}$ associated with each burst we take two approaches: using the gamma-ray energy peak, $E_{peak}$, and the average kinetic isotropic energy, $E_{K,iso}$ to estimate $E_{iso}$. \textit{First}, we assume that our population of \acp{SGRB} follows the isotropic energy ($E_{iso}$) - rest-frame peak energy ($E_z,p$) relationship (see Eq. \ref{eq:eiso-ep}), postulated in Equation 2 of \citet{tsutsui2013possible}. This relationship requires the peak energies of the bursts, $E_p$, which can be obtained by fitting a Band model \citep{band1993batse} to the gamma-ray emission over the duration of the burst. The results of this modelling are usually listed in the public \ac{GBM} catalog \citep{fermi2020cat} and online\footnote{\url{https://heasarc.gsfc.nasa.gov/W3Browse/fermi/fermigbrst.html}}. The compilation of $E_p$ for our \acp{SGRB} sample is listed in Table \ref{table:GRBs}.

\begin{equation}\label{eq:eiso-ep}
    E_{iso} = 10^{52.4\pm0.2} \ erg \ \left( \frac{E_{z,p}}{774.5\ keV}\right) ^{1.6\pm 0.3}
\end{equation}

The energies that result from this transformation are usually larger than the energies derived for previous SGRB afterglows. For this reason, we additionally use the average kinetic isotropic energy, $E_{K,iso}$, presented in \cite{Fong15} as a representative value for $E_{iso}$. Particularly, for this \textit{second} $E_{iso}$ approach, we assume $E_{K,iso} \sim E_{iso} = 2.9\times 10^{51}$ ergs. 

We used the python module \texttt{afterglowpy} \citep{ryan2020gamma} to generate afterglow light-curve templates. Due to the nature of the relativistic jet, we constrained the viewing angle to $\theta < 20^\circ$. We assume a circumburst density of $5.2\times 10^{-3}\ cm^{-3}$, chose a Gaussian jet, and fixed other \texttt{afterglowpy} parameters to standard values: the electron energy distribution index $p=2.43$, as well as the fraction of shock energy imparted to electrons, $\epsilon_E = 0.1$, and to the magnetic field, $\epsilon_B=0.01$. For $E_{iso}$ we used the relation in Eq. \ref{eq:eiso-ep} and the mean $E_{K,iso}$ mentioned in the paragraph above. Additionally for $E_{iso}$ as a function of $E_{z,p}$, we took the gamma-ray $E_{z,p} =  E_p(1+z)$, with the redshift varying for each simulated source. 

We feed \texttt{simsurvey} light-curves generated with \texttt{afterglowpy} assuming the two separate $E_{iso}$ distributions described above. We note that these two approaches are based on conclusions drawn from \textit{Swift} bursts, since the bulk of the SGRB afterglow knowledge comes from \textit{Swift} bursts. We calculated the efficiency as a function of redshift by taking the ratio of sources detected \textit{twice} over the number of generated sources within a redshift volume. We require two detections as our \ac{TOO} strategy relies on at least two data points. 

\begin{figure*}
\begin{center}
\includegraphics[width=0.9\textwidth]{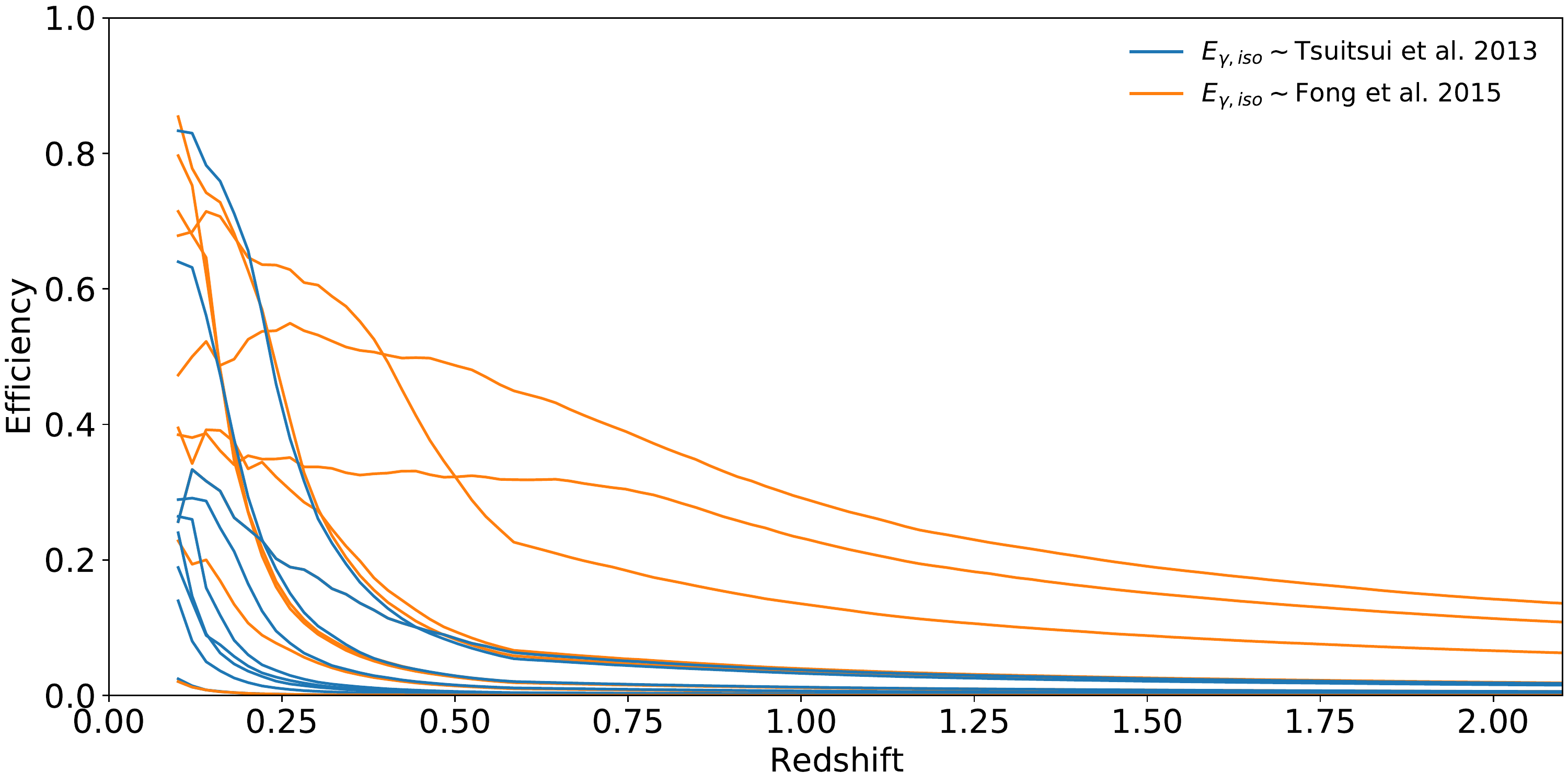}
\includegraphics[width=0.9\textwidth]{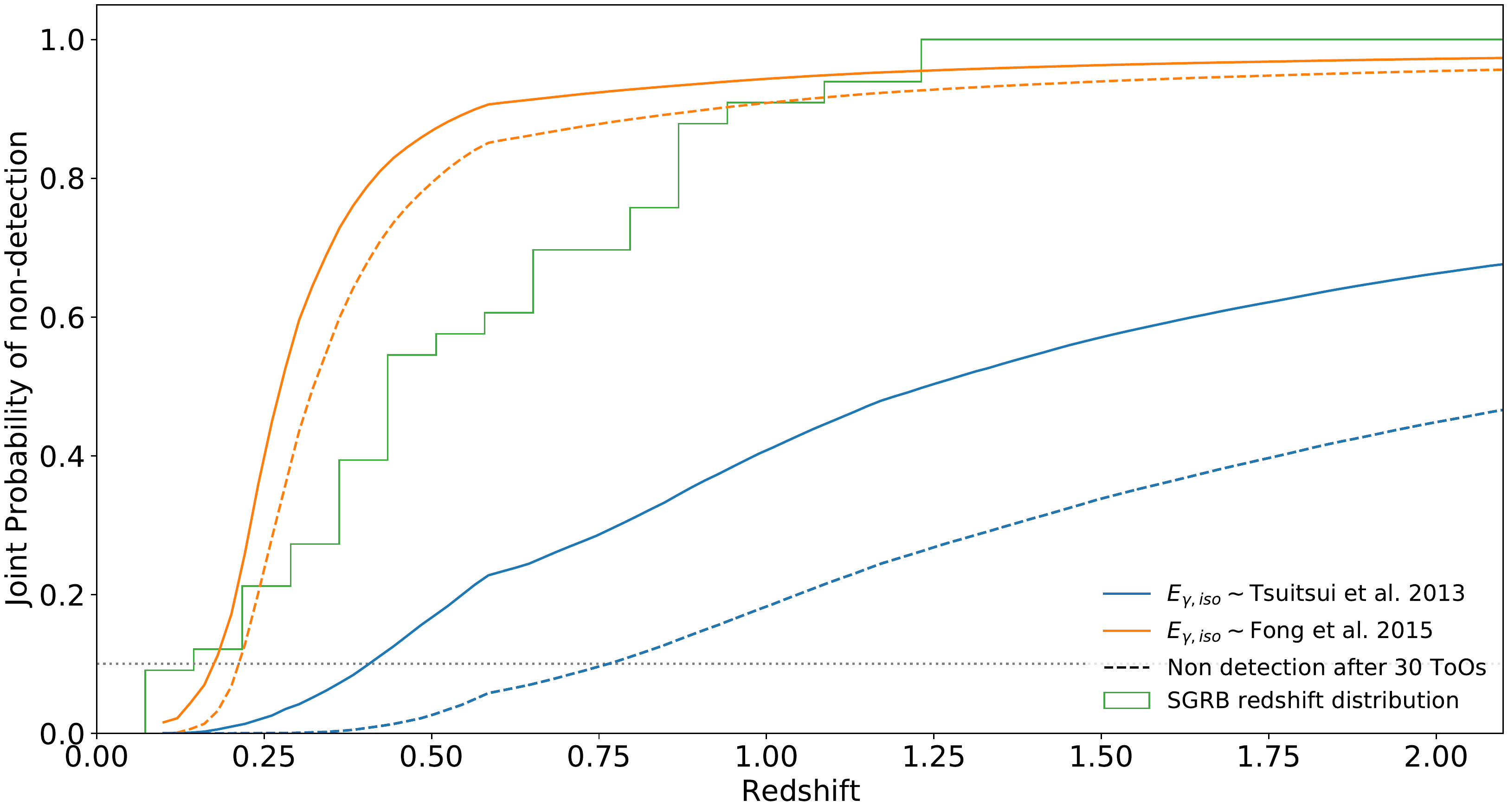}
\end{center}
\caption{(top) The individual efficiency for each SGRB trigger. The blue curves are based on the $E_{iso}$ derived from the Band model $E_p$ and Eq. \ref{eq:eiso-ep}, while the yellow curves are the efficiencies assuming all GRBs have the same $E_{iso}$ as the mean $E_{K,iso}$ from \cite{Fong15}. (bottom) The solid lines represent the joint probability of non-detection using the 9 SGRB triggers with no optical counterparts. We adopt the same color coding as in the top plot, meaning blue for the $E_{iso}$ as a function of $E_{p}$ and yellow for $E_{iso}$  as the mean $E_{K,iso}$ from \cite{Fong15}. The dashed line represent the joint probability of non-detection after 30 \acp{TOO}, assuming an efficiency equal to the median efficiency of the \acp{TOO} presented. We show the cumulative redshift distribution for SGRBs as a green line. The grey dotted line shows the $ CL = 0.9$ level, at which the joint probability of non detection is $1-CL= 0.1$.}
 \label{fig:efficiency_Joint}
 \end{figure*}

The efficiencies vary depending on a few factors. The total coverage and the limiting magnitude of the observations limit the maximum efficiency, which then decays depending on the associated $E_{iso}$. For larger energies, the decay is smoother. In the top panel of Fig. \ref{fig:efficiency_Joint}, we show the efficiencies for the 9 GRBs that had no discovered counterpart. We exclude GRB 200826A as the energies used to model the afterglow follow the SGRB energy distribution, while GRB 200826A was proven to be part of the LGRB population. The energies derived from the \cite{tsutsui2013possible} relationship are larger than the mean $E_{K,iso}$ derived from \cite{Fong15}. This increases the efficiencies at larger redshifts assuming the \cite{tsutsui2013possible} relationship, as the transients are intrinsically more energetic. 

For both of the energies used, we calculate the joint probability of non-detection by taking the product of the SGRB \ac{TOO} efficiencies as a function of redshift. Similar to the analysis in \cite{kasliwal2020kilonova}, we define 
\beq
(1-CL) =  \prod_{i=0}^{N} (1-p_i) 
\eeq
with CL as the credible level and $p_i$ the efficiency of the i$th$ burst as a function of redshift. We show in the bottom panel of Fig. \ref{fig:efficiency_Joint} the result for the afterglows with energies following \cite{tsutsui2013possible} (blue) and \cite{Fong15} (yellow). The lower energies associated with \cite{Fong15} afterglows only allow us to probe the space up to z = 0.16, considering a $CL = 0.9$,  while SGRBs with energies following the $E_{iso}-E_{z,p}$ relationship can be probed as far as z = 0.4. To look into the prospects of the SGRB \ac{TOO} campaign, we model a scenario with 21 additional \ac{TOO} campaigns, each with a median efficiency based on the results presented here. These results are shown as dashed lines in Fig. \ref{fig:efficiency_Joint}, and show that for $E_{iso} \sim E_{K,iso}$, the improvement after thirty \acp{TOO} can only expand our searches (i.e. $CL = 0.9$) up to z = 0.2, while if the GRBs follow the $E_{iso}-E_{z,p}$ relationship, our horizon expands to z = 0.7. 

Finally, when comparing our limits to the redshift distribution of SGRB afterglows found in the literature \citep{Fong15} (green histogram in Fig. \ref{fig:efficiency_Joint}), our searches show that we are probing (and could probe) volumes that contain 10-40\% of the observed afterglows, depending on the $E_{iso}$ assumption.

\section{Proposed follow-up strategy}\label{sec:follow-up-plan}

The current \ac{TOO} strategy aims for two consecutive exposures in two different filters, prioritizing the color of the source as the main avenue to discriminate between sources. This helps confirming the nature of the transient as an extragalactic source. In some cases, it can lead to problems as the source might not be detected at shorter wavelengths, due to either the extinction along the line of sight or its intrinsically fainter brightness. If there is no second detection at shorter wavelengths, there is the risk of ignoring a potential counterpart as a single detection can be confused as a slow moving object or an artifact. The standard strategy considers a second night of ZTF observations in the same two filters, to measure the magnitude and color evolution. However, a number of sources did not have a second detection in the same filter after the second night, impeding the measurement of the decline rate. For these two reasons, for afterglow searches with ZTF (and possibly other instruments with similar limiting magnitudes), it is more informative to observe the region at least twice in the same filter during the first night. By separating the two same-filter epochs by at least $2 \sigma \times 24/\alpha$, where $\sigma$ is the typical error of the observations and $\alpha$ is the power-law index of the afterglow decline, we can possibly measure the decay rate of sources, or at least set a lower limit for $\alpha$. For ZTF, two epochs separated by 6 hours would suffice for afterglows with a typical $\alpha \sim 1$, assuming $\sigma = 0.12$. 

This scenario is unlikely to happen often, as it requires that the region is visible during the entire night and that the night is long enough to allow for two visits separated by a number of hours. In any case, the standard \ac{TOO} strategy for the second night of observation (two visits in two different filters) should help determine the color and magnitude evolution.
 
\begin{figure}[!ht]
 \includegraphics[width=0.48\textwidth]{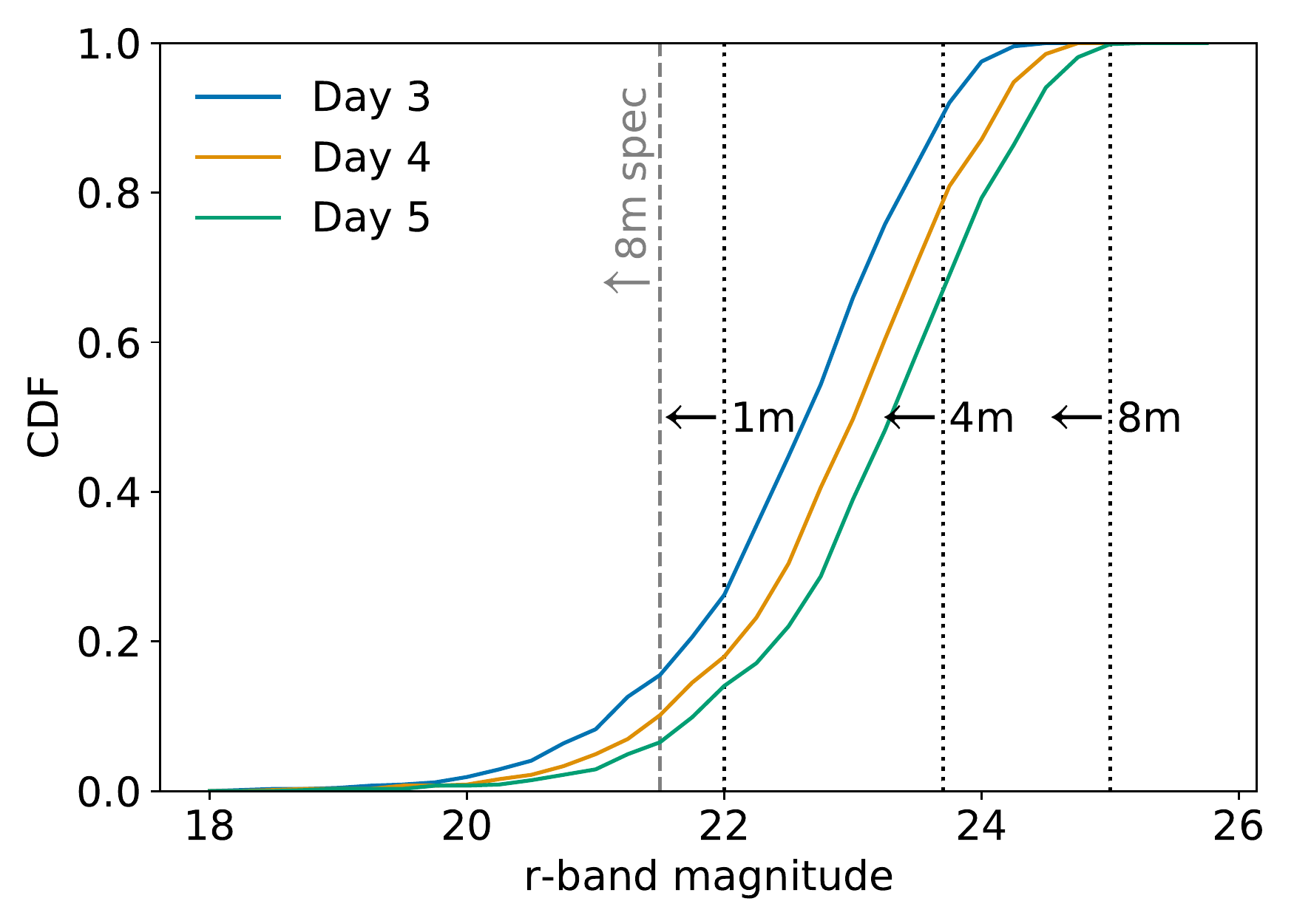}
 \caption{The magnitude cumulative distribution of the sources detected using \texttt{simsurvey} as a function of the days after the burst. This distribution contains all the sources detected up to z$=2$. The photometric and spectroscopic limits of different facilities are shown as dotted vertical lines. }
 \label{fig:follow-up}
\end{figure}

 For the third day of follow-up, there will be two kinds of candidates: (a) confirmed fast fading transients, and (b) transients with unconstrained evolution, that likely only have data for the first night. For (a) it is important to get spectra as soon as possible before the transients fade below the spectroscopic limits. Ideally, observations in other wavelengths should be triggered to cement the classification and begin the characterization of the transient. For candidates in situation (b), the fast evolution of the transients requires the use of larger facilities. From our experience, this is feasible as only a handful of candidates will fall in this category. In both cases, (a) and (b), photometric follow-up using facilities different than ZTF are needed, as any afterglow detected by ZTF will likely not be detectable three days after the burst. In Fig. \ref{fig:follow-up} we show the magnitude distribution of all the transients that \texttt{simsurvey} detected, independent of redshift, as a function of how many days passed after the burst. This figure illustrates the need for other telescopes to monitor the evolution of the transient, as for example, only $\sim$30\% of the transients that we can detect with ZTF will be brighter than $r>22$ mag. Additionally, Fig.~\ref{fig:follow-up} shows that spectroscopy of the sources becomes harder after day 2, as only 20\% of the detected transients will be brighter than $r=21.5$ mag. 
 
 Since spectroscopic data will be challenging to acquire for faint sources, the panchromatic follow-up, from radio to x-rays, will help to confirm the classification of the transient.  

\section{Conclusions}\label{sec:conclusion}

 During a period of $\sim 2$ years, a systematic, extended and deep search for the optical counterparts to \textit{Fermi}-\ac{GBM} \acp{SGRB} has been performed employing the Zwicky Transient Facility. The \ac{ZTF} observations of 10 events followed-up are listed in Table \ref{table:GCNs} and no optical counterpart has yet been associated to a compact binary coalescence. However, our \ac{TOO} strategy led to the discovery of the optical counterpart to GRB 200826A, which was ultimately revealed as the shortest-duration LGRB found to date \citep{ahumada2021}.
 
This experiment complements previous studies \citep{SiCe2013,SiKa2015,CoAh19sgrb..131d8001C}, and demonstrates the feasibility of studying the large sky areas derived from \textit{Fermi} \ac{GBM} by exploiting the wide field of view of \ac{ZTF}. The average coverage was  $\sim 60\%$ of the localization regions, corresponding to $\sim 950$ deg$^2$. The average amount of alerts in the targeted regions of the sky was over 20000, and we were able to reduce this figure to no more than 20 candidates per trigger. Thanks to the high cadence of \ac{ZTF} we were able to achieve a median reduction in alerts of 0.03\%. The effectiveness of the filtering criteria is comparable with the median reduction reached in \cite{SiKa2015}, even when the areas covered are almost orders of magnitude larger. The \ac{IPTF} search for the optical counterparts to the long gamma-ray burst GRB 130702A covered 71 deg$^2$ and yielded 43 candidates \citep{SiCe2013}.

This campaign has utilized \ac{ZTF} capabilities to rapidly follow-up \ac{SGRB} trigger, which has allowed us to explore the magnitude space and set constraints on \acp{SGRB} events. The average depth for \ac{ZTF} 300s exposures is $r\sim20.8$ which has allowed us to look for SGRB afterglows and GW170817-like KNe. From Fig. \ref{fig:follow-up}, it can be seen that future follow-ups would benefit both from a more rapid response and longer exposures.

By using computational tools like \texttt{afterglowpy} and \texttt{simsurvey}, we have quantified the efficiency of our \ac{TOO} triggers. The ZTF efficiency drops quickly as the transient is located at further distances, and the magnitude limits only allow for detections up to z = 0.4, for energies following the \cite{tsutsui2013possible} relation and z = 0.16 for bursts with energies equal to the mean $E_{iso}$ found by \cite{Fong15}, for a $CL = 0.9$. Furthermore, when repeating the experiment 21 times (to complete 30 \acp{TOO}) and assuming a median efficiency $p_{med}$ for each new event, the horizons of our searches increase to z = 0.2 and 0.72 respectively. 

Additionally, our simulations show that ZTF is no longer effective at following-up afterglows after three days following the burst. The fast fading nature of these transients requires deeper observations, and spectroscopic and panchromatic observations are helpful to reveal the nature of the candidates. Ideally, at least two observations in the same filter should be taken during the first night of observation, as afterglows and KNe fade extremely rapidly and they might not be observable 48\,hrs after the burst. With this strategy we can hope to find another counterpart.

\begin{acknowledgements}
%ZTF 
Based on observations obtained with the Samuel Oschin Telescope 48-inch and the 60-inch Telescope at the Palomar Observatory as part of the Zwicky Transient Facility project. ZTF is supported by the National Science Foundation under Grants No. AST-1440341 and AST-2034437 and a collaboration including current partners Caltech, IPAC, the Weizmann Institute for Science, the Oskar Klein Center at Stockholm University, the University of Maryland, Deutsches Elektronen-Synchrotron and Humboldt University, the TANGO Consortium of Taiwan, the University of Wisconsin at Milwaukee, Trinity College Dublin, Lawrence Livermore National Laboratories, IN2P3, University of Warwick, Ruhr University Bochum, Northwestern University and former partners the University of Washington, Los Alamos National Laboratories, and Lawrence Berkeley National Laboratories. Operations are conducted by COO, IPAC, and UW.
%growth
This work was supported by the GROWTH (Global Relay of Observatories Watching Transients Happen) project funded by the National Science Foundation under PIRE Grant No 1545949. GROWTH is a collaborative project among California Institute of Technology (USA), University of Maryland College Park (USA), University of Wisconsin Milwaukee (USA), Texas Tech University (USA), San Diego State University (USA), University of Washington (USA), Los Alamos National Laboratory (USA), Tokyo Institute of Technology (Japan), National Central University (Taiwan), Indian Institute of Astrophysics (India), Indian Institute of Technology Bombay (India), Weizmann Institute of Science (Israel), The Oskar Klein Centre at Stockholm University (Sweden), Humboldt University (Germany), Liverpool John Moores University (UK) and University of Sydney (Australia).
%lsst dsfp
TA and HK thank the LSSTC Data Science Fellowship Program, which is funded by LSSTC, NSF Cybertraining Grant \#1829740, the Brinson Foundation, and the Moore Foundation; their participation in the program has benefited this work. 
% SRM and JC acknowledge support from the NSF grant NSF PHY \#1912649. We are grateful for computational resources provided by the Leonard E Parker Center for Gravitation, Cosmology and Astrophysics at the University of Wisconsin-Milwaukee. 
MMK acknowledges generous support from the David and Lucille Packard Foundation. 
M.W.C. acknowledges support from the National Science Foundation with grant Nos. PHY-2010970 and OAC-2117997. 
S.A. acknowledges support from the GROWTH PIRE Grant No 1545949. 
ASC acknowledges support from the G.R.E.A.T research environment, funded by {\em Vetenskapsr\aa det},  the Swedish Research Council, project number 2016-06012.
M.B. acknowledges support from the Swedish Research Council (Reg. no. 2020-03330).
S.R. acknowledges support by the Helmholtz Weizmann Research School on Multimessenger Astronomy, funded through the Initiative and Networking Fund of the Helmholtz Association, DESY, the Weizmann Institute, the Humboldt University of Berlin, and the University of Potsdam.
ECK acknowledges support from the G.R.E.A.T research environment funded by {\em Vetenskapsr\aa det}, the Swedish Research Council, under project number 2016-06012, and support from The Wenner-Gren Foundations.
% T.A. and H.K. are LSSTC Data Science Fellows and thank the LSSTC Data Science Fellowship Program, which is funded by LSSTC, NSF Cybertraining grant No. 1829740, the Brinson Foundation and the Moore Foundation; their participation in the program has benefited this work. 
P.R. acknowledges the support received from the Agence Nationale de la Recherche of the French government through the program ‘‘Investissements d’Avenir’’ (16-IDEX-0001 CAP 20-25)
%NASA
The material is based on work supported by NASA under award No. 80GSFC17M0002.
% gemini
Based on observations obtained at the international Gemini Observatory, a program of NSF’s NOIRLab, which is managed by the Association of Universities for Research in Astronomy (AURA) under a cooperative agreement with the National Science Foundation on behalf of the Gemini Observatory partnership: the National Science Foundation (United States), National Research Council (Canada), Agencia Nacional de Investigación y Desarrollo (Chile), Ministerio de Ciencia, Tecnología e Innovación (Argentina), Ministério da Ciência, Tecnologia, Inovações e Comunicações (Brazil), and Korea Astronomy and Space Science Institute (Republic of Korea). The observations were obtained as part of Gemini Director’s Discretionary Program GN-2021A-Q-102. The Gemini data were processed using DRAGONS (Data Reduction for Astronomy from Gemini Observatory North and South). This work was enabled by observations made from the Gemini North telescope, located within the Maunakea Science Reserve and adjacent to the summit of Maunakea. We are grateful for the privilege of observing the Universe from a place that is unique in both its astronomical quality and its cultural significance. 
% ZTF FP
The ZTF forced-photometry service was funded under the Heising-Simons Foundation grant No. 12540303 (PI: Graham). 
% LDT
These results also made use of Lowell Observatory’s Lowell Discovery Telescope (LDT), formerly the Discovery Channel Telescope. Lowell operates the LDT in partnership with Boston University, Northern Arizona University, the University of Maryland and the University of Toledo. Partial support of the LDT was provided by Discovery Communications. LMI was built by Lowell Observatory using funds from the National Science Foundation (AST-1005313). 
% LT
The Liverpool Telescope is operated on the island of La Palma by Liverpool John Moores University in the Spanish Observatorio del Roque de los Muchachos of the Instituto de Astrofisica de Canarias with financial support from the UK Science and Technology Facilities Council.
%SEDM
SED Machine is based upon work supported by the National Science Foundation under Grant No. 1106171.
%GIT
GIT is a 70-cm telescope with a 0\fdg7 field of view, set up by the Indian Institute of Astrophysics (IIA) and the Indian Institute of Technology Bombay (IITB) with funding from DST-SERB and IUSSTF. It is located at the Indian Astronomical Observatory, operated by IIA. We acknowledge funding by the IITB alumni batch of 1994, which partially supports operations of the telescope. Telescope technical details are available at \url{https://sites.google.com/view/growthindia/}.

\end{acknowledgements}

\software{{\tt ipython} \citep{ipython}, {\tt jupyter} \citep{jupyter}, {\tt matplotlib} \citep{matplotlib}, {\tt python} \citep{python3}, {\tt NumPy} \citep{numpy}, {\tt afterglowpy} \citep{ryan2020gamma}, {\tt simsurvey} \citep{feindt2019simsurvey}}%,  {\tt ligo.skymap} \citep{ligoskymap}}

\facilities{\textit{Fermi}-GBM, ZTF/PO:1.2m, P60, P200, KPED, LCOGT, Gemini, LDT, Keck, LT, GIT }

\bibliographystyle{aasjournal}
\bibliography{references}

%% Candidates and tables

\begin{deluxetable*}{ccccccc}
\tablecaption{\label{table:GCNs} Summary of the \ac{ZTF} \ac{TOO} triggers. We list the area covered with ZTF, as well as the corresponding credible region (C.R.) of the GBM map. We shown our time delay between the burst and the start of ZTF observations. For each trigger, we list the exposure time for night 1 and night 2, along with the filter sequence in parenthesis. The last two columns show the median $r-$band 5$\sigma$ limit and the number of objects followed-up with other facilities. }
 \tablehead{\colhead{GRB} & \colhead{Area covered} & \colhead{{\begin{tabular}{@{}c@{}}C.R. \\ covered\end{tabular}}} &  \colhead{{\begin{tabular}{@{}c@{}}Time delay \\ in triggering ZTF\end{tabular}}} &   \colhead{{\begin{tabular}{@{}c@{}}Exposure time \\ (sequence)\end{tabular}} } & \colhead{$r-$band 5$\sigma$ limit} & \colhead{{\begin{tabular}{@{}c@{}}Objects \\ followed-up\end{tabular}} }}
 \startdata
GRB 180523B &2900 $deg^2$& 60\% & 9.1h & 60s(rgr), 90s(rgr) &  r $>$ 20.3 mag&14\\
GRB 180626C &275 $deg^2$&87\% & 1.5h & 120s(rgr), 240s(grg)  &r $>$ 20.9 mag& 1\\
GRB 180715B &254 $deg^2$&37\% & 10.3h & 180s(rgr), 240s(rg) &r $>$ 21.4 mag&14\\
GRB 180728B &334 $deg^2$& 76\% & 31h & 180s(rgr), 180s(rgr) &r $>$ 18.7 mag&7\\
GRB 180913A &546 $deg^2$& 53\% & 8.3h & 180s(grg), 300s(grg) &r $>$ 22.2 mag&12\\
GRB 181126B &1400 $deg^2$& 66\% & 1.3h & 180s(rr), 300s (r) &r $>$ 20.5 mag&11 \\
GRB 200514B &519 $deg^2$& 49\% & 0.9h & 300s(gr) &r $>$ 22.2 mag&14 \\
GRB 201130A &400 $deg^2$& 75\% & 7h& 300s(grg),300s(gr) &r $>$ 20.3 mag&0 \\
GRB 210510A &1105 $deg^2$& 84\% & 10h& 180(gr),240(r) &r $>$ 22.1 mag&1 \\
\enddata
\end{deluxetable*}

%%%% new long table
\startlongtable
\begin{deluxetable*}{ccccccc}
\tablecaption{ \label{table:candidates_all} Follow-up table of the candidates identified for GRB 180523B \citep{gcn22739}, GRB 180626C \citep{gcn22871}, GRB 180715B \citep{gcn22969}, GRB 180913A \citep{gcn23324}, GRB 181126B \citep{gcn23515}, GRB 200514B \citep{gcn27737,gcn27745}, and GRB 210510A \citep{gcn30005}. The spectroscopic (s) or photometric (p) redshifts of the respective host galaxies are listed as well. The photometric slow evolution of some candidates was used as a rejection criteria when the object presents a variation on its magnitude smaller than 0.3 mag/day.}
\tablehead{\colhead{GRB trigger} & \colhead{ZTF Name} & \colhead{RA} &  \colhead{Dec} &   \colhead{{\begin{tabular}{@{}c@{}}Discovery \\ magnitude \end{tabular}} } & \colhead{Redshift} & \colhead{Rejection criteria} }
\startdata
GRB 180523B &ZTF18aawozzj  & 12:31:09.02 & +57:35:01.8 & g = 20.20 & (s) 0.095 & SN Ia-91T P200 \\ 
&ZTF18aawnbgg  & 10:40:54.05 & +23:44:43.3 & r = 19.80 & (s) 0.135 & SN Ia P200 \\ 
&ZTF18aawmvbj  & 10:12:41.17 & +21:24:55.5 & r = 19.75 & (s) 0.14 & SN Ia P200 \\ 
&ZTF18aawcwsx  & 10:40:33.46 & +47:02:24.4 & r = 19.84 & (s) 0.09 & SN Ia-91T P60 \\ 
&ZTF18aawnbkw  & 10:38:47.66 & +26:18:51.8 & r = 19.91 & (p) 0.31 & slow SDSS\\ 
&ZTF18aawmqwo  & 09:52:06.90 & +47:18:34.8 & r = 19.98 & (p) 0.04 & slow SDSS \\ 
&ZTF18aawmkik  & 08:51:11.45 & +13:13:16.7 & r = 19.04 & (p) 0.52 & slow SDSS \\ 
&ZTF18aawnmlm  & 11:03:11.38 & +42:07:29.9 & r = 20.12 & orphan & slow flat in 7 days \\ 
&ZTF18aauhzav  & 10:59:29.32 & +44:10:02.7 & r = 19.97 & (s) 0.05 & slow 2MASX \\ 
&ZTF18aavrhqs  & 11:58:09.57 & +63:45:34.6 & r = 19.99 & orphan & slow  \\ 
&ZTF18aawmwwk  & 10:35:26.51 & +65:22:34.3 & r = 19.99 & (p) 0.18 & slow SDSS \\ 
&ZTF18aawwbwm  & 08:16:44.98 & +35:34:13.1 & r = 19.79 & (p) 0.15 & slow SDSS \\ 
&ZTF18aawmjru  & 08:39:11.39 & +44:01:53.6 & r = 18.43 & (p) 0.44 & slow SDSS \\ 
&ZTF18aawmigr  & 08:48:01.76 & +29:13:51.9 & r = 19.63 & (s) 0.1 & slow 2MASX \\ \hline
GRB 180626C & ZTF18aauebur  & 19:48:49.10 & +46:30:36.1 & r = 18.85 & stellar & CV multiple previous bursts \\ \hline
GRB 180715B &ZTF18aamwzlv & 13:06:44.59 & +68:59:52.9 & r = 18.50 & (s) 0.1 & slow  \\ 
&ZTF18abhbevp & 14:21:00.83 & +72:11:43.8 & g = 20.63 & -- & slow \\ 
&ZTF18abhbpkm & 16:02:36.78 & +70:47:05.1 & g = 21.24 & -- & slow\\ 
&ZTF18abhhjyd & 13:02:32.07 & +75:16:49.4 & g = 20.43 & -- & AGN Milliquas match \\ 
&ZTF18abhbgan & 15:43:18.86 & +72:05:24.8 & g = 21.22 & orphan & slow \\ 
&ZTF18abhbfoi & 13:24:34.01 & +70:56:47.5 & g = 21.12 & (s )1.2 & AGN Milliquas and PS1\\ 
&ZTF18abhbcjy & 14:20:50.39 & +73:25:40.5 & g = 20.78 & -- & slow \\ 
&ZTF18abhaogg & 13:42:45.47 & +74:19:38.3 & r = 20.38 & orphan & slow \\ 
&ZTF18abhbamj & 15:26:58.78 & +72:02:17.8 & r = 21.27 & orphan & slow \\ 
&ZTF18abhawjn & 13:31:27.33 & +66:46:45.4 & g = 20.69 & (s) 0.4 & AGN Milliquas \\ 
&ZTF18abharzk & 13:41:09.05 & +70:43:06.8 & r = 21.30 & -- & slow \\ 
&ZTF18abhbckn & 12:49:53.85 & +73:02:00.5 & r = 20.93 & (s) 0.00541 & slow CLU \\ 
&ZTF18abhbfqf & 13:16:00.24 & +69:37:24.1 & r = 19.80 & (s) 0.11 & SN Ia-91T P200 \\ 
&ZTF18aauhpyb & 13:21:45.49 & +70:55:59.8 & g = 19.67 & stellar & CV multiple bursts P60\\ \hline
GRB 180913A&ZTF18abvzgms & 23:37:50.57 & +47:53:21.2 & g = 21.29 & (p) 0.35 & flat evolution SDSS\\ 
&ZTF18abwiios & 23:12:14.06 & +39:27:50.6 & g = 22.04 & -- & flat evolution \\   
&ZTF18abvzfgy & 23:16:15.20 & +43:31:59.3 & g = 20.98 & (s) 0.04 & SN Ic LDT  \\ 
&ZTF18abvzjwk & 	22:30:32.49 & +39:50:14.6 & g = 21.70 & -- & flat evolution  \\ 
&ZTF18abvwhkl & 	23:05:44.17 & +45:32:34.8 & r = 21.44 & -- & flat evolution 3 points\\ 
&ZTF18abvucnv & 	22:31:31.96 & +39:30:03.7 & r = 21.15 & stellar & Star flare \\  
&ZTF18abwiitm & 	23:15:27.61 & +39:57:10.5 & g = 21.71 & -- & slow AGN WISE \\ 
&ZTF18abvubdm & 	22:58:28.45 & +47:06:03.8 & g = 21.01 & -- & slow evolution nice lc \\ 
&ZTF18abvzsld & 	00:15:57.12 & +49:28:51.0 & g = 21.50 & Stellar & flat evolution \\ 
&ZTF18abwiivr & 	22:52:15.80 & +37:22:29.4 & g = 21.73 & Stellar & slow evolution \\ 
&ZTF18abvzmtm & 	23:55:13.07 & +48:21:37.8 & g = 21.65 & -- & slow \\ \hline
GRB 181126B&ZTF18achtkfy & 	06:54:02.63 & +37:04:28.6 & g = 19.69 & orphan &  slow\\ 
&ZTF18achflqs & 	04:41:09.49 & +23:53:24.9 & r = 20.20 & (p) 0.38 &  flat evolution SDSS\\ 
&ZTF18acrkcxa & 	04:55:02.52 & +22:40:43.4 & r = 20.85 & Stellar &  Flare Keck LRIS\\ 
&ZTF18acrkkpc & 	06:23:15.56 & +10:19:22.6 & r = 20.17 & (s) 0.061 & SN II Keck LRIS\\ 
&ZTF18aadwfrc & 	06:17:18.02 & +50:29:03.3 & r = 19.65 & (s) 0.04 &  SN Ia-02cx Keck LRIS \\ 
&ZTF18acrfond & 	03:59:26.95 & +24:35:20.4 & r = 10.13 & (s) 0.117 & SN Ia Keck LRIS \\ 
&ZTF18acrfymv & 	06:18:01.18 & +44:10:52.7 & g = 20.82 & (s) 0.072 & SN Ic-BL Keck LRIS \\ 
&ZTF18acptgzz & 	04:33:32.45 & -01:38:51.1 & r = 19.56 & (s) 0.096 & SN Ia Keck LRIS  \\ 
&ZTF18acbyrll & 	05:55:28.67 & +29:28:20.3 & r = 19.34 & -- &  slow evolution \\ 
&ZTF18acrewzd & 	04:41:17.29& -01:46:07.5 & g = 20.74&  (s) 0.13 &  SN Ia Keck LRIS\\ \hline
GRB 200514B &ZTF20aazpphd & 242.7149675 & +27.1616870 & r = 19.6 & --& slow\\
 &ZTF20aazppnv & 238.1438691 & +25.5764946 & r = 21.1 & (p) 0.17 &  slow\\
 &ZTF20aazprjq & 233.5213585 & +43.3298714 & r = 21.3  &(p) 0.23& slow \\
 &ZTF20aazptlp & 229.007524  & +48.774925  & r = 21.5  &(p) 0.40& slow \\
 &ZTF20aazptnn & 237.2967278 & +47.271954  & r = 21.6  &(p) 0.26& slow \\
 &ZTF20aazpnst & 254.0989833 & +34.4655542 & r = 22.0  &(p) 0.19& slow \\
 &ZTF20aazpofi & 236.929525  & +46.9809542 & r = 21.5  &(p) 0.46& slow \\
 &ZTF20aazplwp & 2734.0167814 & 41.1672761 & r = 21.6 & -- &  slow\\
 &ZTF20aazqlgx & 2746.0908608 & 34.6259478 & r = 22.3 & (p) 0.35 & slow \\
 &ZTF20aazphye & 2755.6577428 & 41.7013160 & r = 21.6 & (p) 0.26 & slow \\
 &ZTF20aazpnxd & 2755.931646  & 48.3862806 & r = 21.6 & --& slow \\
 &ZTF20aazpkri & 2740.7324792 & 48.5554957 & r = 21.3 & --& slow \\
 &ZTF20aazqndp & 2737.8212032 & 50.4933039 & r = 22.1 & (s) 0.03 &  slow\\
 &ZTF20aazqpps & 2752.2388065 & 41.3097433 & r = 21.6 & (s) 0.2 &  slow\\\hline 
GRB 210510A & ZTF21abaytuk & 13:48:49.89 & +35:32:13.05 &  g  = 21.76 & (s) 0.8970 & AGN Keck LRIS\\
\enddata
\end{deluxetable*}

\end{document}

%% file: acronyms.tex
\providecommand{\acrolowercase}[1]{\lowercase{#1}}

\begin{acronym}
\acro{2D}[2D]{two\nobreakdashes-dimensional}
\acro{2+1D}[2+1D]{2+1\nobreakdashes--dimensional}
\acro{2MRS}[2MRS]{2MASS Redshift Survey}
\acro{3D}[3D]{three\nobreakdashes-dimensional}
\acro{2MASS}[2MASS]{Two Micron All Sky Survey}
\acro{AdVirgo}[AdVirgo]{Advanced Virgo}
\acro{AMI}[AMI]{Arcminute Microkelvin Imager}
\acro{AGN}[AGN]{active galactic nucleus}
\acroplural{AGN}[AGN]{active galactic nuclei}
\acro{aLIGO}[aLIGO]{Advanced \acs{LIGO}}
\acro{ASKAP}[ASKAP]{Australian \acl{SKA} Pathfinder}
\acro{ATCA}[ATCA]{Australia Telescope Compact Array}
\acro{ATLAS}[ATLAS]{Asteroid Terrestrial-impact Last Alert System}
\acro{BAT}[BAT]{Burst Alert Telescope\acroextra{ (instrument on \emph{Swift})}}
\acro{BATSE}[BATSE]{Burst and Transient Source Experiment\acroextra{ (instrument on \acs{CGRO})}}
\acro{BAYESTAR}[BAYESTAR]{BAYESian TriAngulation and Rapid localization}
\acro{BBH}[BBH]{binary black hole}
\acro{BHBH}[BHBH]{\acl{BH}\nobreakdashes--\acl{BH}}
\acro{BH}[BH]{black hole}
\acro{BNS}[BNS]{binary neutron star}
\acro{CARMA}[CARMA]{Combined Array for Research in Millimeter\nobreakdashes-wave Astronomy}
\acro{CASA}[CASA]{Common Astronomy Software Applications}
\acro{CBCG}[CBCG]{Compact Binary Coalescence Galaxy}
\acro{CFH12k}[CFH12k]{Canada--France--Hawaii $12\,288 \times 8\,192$ pixel CCD mosaic\acroextra{ (instrument formerly on the Canada--France--Hawaii Telescope, now on the \ac{P48})}}
\acro{CLU}[CLU]{Census of the Local Universe}
\acro{CRTS}[CRTS]{Catalina Real-time Transient Survey}
\acro{CTIO}[CTIO]{Cerro Tololo Inter-American Observatory}
\acro{CBC}[CBC]{compact binary coalescence}
\acro{CCD}[CCD]{charge coupled device}
\acro{CDF}[CDF]{cumulative distribution function}
\acro{CGRO}[CGRO]{Compton Gamma Ray Observatory}
\acro{CMB}[CMB]{cosmic microwave background}
\acro{CRLB}[CRLB]{Cram\'{e}r\nobreakdashes--Rao lower bound}
\acro{CV}[CV]{Cataclysmic Variable}
\acro{cWB}[\acrolowercase{c}WB]{Coherent WaveBurst}
\acro{DASWG}[DASWG]{Data Analysis Software Working Group}
\acro{DBSP}[DBSP]{Double Spectrograph\acroextra{ (instrument on \acs{P200})}}
\acro{DCT}[DCT]{Discovery Channel Telescope}
\acro{DECAM}[DECam]{Dark Energy Camera\acroextra{ (instrument on the Blanco 4\nobreakdashes-m telescope at \acs{CTIO})}}
\acro{DES}[DES]{Dark Energy Survey}
\acro{DFT}[DFT]{discrete Fourier transform}
\acro{EM}[EM]{electromagnetic}
\acro{ER8}[ER8]{eighth engineering run}
\acro{FD}[FD]{frequency domain}
\acro{FAR}[FAR]{false alarm rate}
\acro{FFT}[FFT]{fast Fourier transform}
\acro{FIR}[FIR]{finite impulse response}
\acro{FITS}[FITS]{Flexible Image Transport System}
\acro{F2}[F2]{FLAMINGOS\nobreakdashes-2}
\acro{FLOPS}[FLOPS]{floating point operations per second}
\acro{FOV}[FOV]{field of view}
\acroplural{FOV}[FOV\acrolowercase{s}]{fields of view}
\acro{FP}[FP]{forced photometry}
\acro{FTN}[FTN]{Faulkes Telescope North}
\acro{FWHM}[FWHM]{full width at half-maximum}
\acro{GALEX}[GALEX]{Galaxy Evolution Explorer}
\acro{GBM}[GBM]{Gamma-ray Burst Monitor\acroextra{ (instrument on \emph{Fermi})}}
\acro{GCN}[GCN]{Gamma-ray Coordinates Network}
\acro{GIT}[GIT]{GROWTH India telescope }
\acro{GLADE}[GLADE]{Galaxy List for the Advanced Detector Era}
\acro{GMOS}[GMOS]{Gemini Multi-object Spectrograph\acroextra{ (instrument on the Gemini telescopes)}}
\acro{GOTO}[GOTO]{Gravitational-Wave Optical Transient Observer}
\acro{GRB}[GRB]{gamma-ray burst}
\acro{GROWTH}[GROWTH]{Global Relay of Observatories Watching Transients Happen}
\acro{GSC}[GSC]{Gas Slit Camera}
\acro{GSL}[GSL]{GNU Scientific Library}
\acro{GTC}[GTC]{Gran Telescopio Canarias}
\acro{GW}[GW]{gravitational wave}
\acro{GWGC}[GWGC]{Gravitational Wave Galaxy Catalogue}
\acro{HAWC}[HAWC]{High\nobreakdashes-Altitude Water \v{C}erenkov Gamma\nobreakdashes-Ray Observatory}
\acro{HCT}[HCT]{Himalayan Chandra Telescope}
\acro{HEALPix}[HEALP\acrolowercase{ix}]{Hierarchical Equal Area isoLatitude Pixelization}
\acro{HEASARC}[HEASARC]{High Energy Astrophysics Science Archive Research Center}
\acro{HETE}[HETE]{High Energy Transient Explorer}
\acro{HFOSC}[HFOSC]{Himalaya Faint Object Spectrograph and Camera\acroextra{ (instrument on \acs{HCT})}}
\acro{HMXB}[HMXB]{high\nobreakdashes-mass X\nobreakdashes-ray binary}
\acroplural{HMXB}[HMXB\acrolowercase{s}]{high\nobreakdashes-mass X\nobreakdashes-ray binaries}
\acro{HSC}[HSC]{Hyper Suprime\nobreakdashes-Cam\acroextra{ (instrument on the 8.2\nobreakdashes-m Subaru telescope)}}
\acro{IACT}[IACT]{imaging atmospheric \v{C}erenkov telescope}
\acro{IIR}[IIR]{infinite impulse response}
\acro{IMACS}[IMACS]{Inamori-Magellan Areal Camera \& Spectrograph\acroextra{ (instrument on the Magellan Baade telescope)}}
\acro{IMR}[IMR]{inspiral-merger-ringdown}
\acro{IPAC}[IPAC]{Infrared Processing and Analysis Center}
\acro{IPN}[IPN]{InterPlanetary Network}
\acro{IPTF}[\acrolowercase{i}PTF]{intermediate \acl{PTF}}
\acro{IRAC}[IRAC]{Infrared Array Camera}
\acro{ISM}[ISM]{interstellar medium}
\acro{ISS}[ISS]{International Space Station}
\acro{KAGRA}[KAGRA]{KAmioka GRAvitational\nobreakdashes-wave observatory}
\acro{KDE}[KDE]{kernel density estimator}
\acro{KN}[KN]{kilonova}
\acroplural{KN}[KNe]{kilonovae}
\acro{KPED}[KPED]{Kitt Peak Electron multiplying CCD Demonstrator}
\acro{LAT}[LAT]{Large Area Telescope}
\acro{LCOGT}[LCOGT]{Las Cumbres Observatory Global Telescope}
\acro{LDT}[LDT]{Lowell Discovery Telescope}
\acro{LGRB}[LGRB]{long \acl{GRB}}
\acro{LHO}[LHO]{\ac{LIGO} Hanford Observatory}
\acro{LIB}[LIB]{LALInference Burst}
\acro{LIGO}[LIGO]{Laser Interferometer \acs{GW} Observatory}
\acro{llGRB}[\acrolowercase{ll}GRB]{low\nobreakdashes-luminosity \ac{GRB}}
\acro{LLOID}[LLOID]{Low Latency Online Inspiral Detection}
\acro{LLO}[LLO]{\ac{LIGO} Livingston Observatory}
\acro{LMI}[LMI]{Large Monolithic Imager\acroextra{ (instrument on \ac{DCT})}}
\acro{LOFAR}[LOFAR]{Low Frequency Array}
\acro{LOS}[LOS]{line of sight}
\acroplural{LOS}[LOSs]{lines of sight}
\acro{LMC}[LMC]{Large Magellanic Cloud}
\acro{LRIS}[LRIS]{Low Resolution Imaging Spectrograph}
\acro{LSB}[LSB]{long, soft burst}
\acro{LSC}[LSC]{\acs{LIGO} Scientific Collaboration}
\acro{LSO}[LSO]{last stable orbit}
\acro{LSST}[LSST]{Large Synoptic Survey Telescope}
\acro{LT}[LT]{Liverpool Telescope}
\acro{LTI}[LTI]{linear time invariant}
\acro{LVC}[LVC]{the LIGO-Virgo Collaboration}
\acro{MAP}[MAP]{maximum a posteriori}
\acro{MBTA}[MBTA]{Multi-Band Template Analysis}
\acro{MCMC}[MCMC]{Markov chain Monte Carlo}
\acro{MLE}[MLE]{\ac{ML} estimator}
\acro{ML}[ML]{maximum likelihood}
\acro{MPC}[MPC]{Minor Planet Center}
\acro{MOU}[MOU]{memorandum of understanding}
\acroplural{MOU}[MOUs]{memoranda of understanding}
\acro{MWA}[MWA]{Murchison Widefield Array}
\acro{NED}[NED]{NASA/IPAC Extragalactic Database}
\acro{NIR}[NIR]{near infrared}
\acro{NSBH}[NSBH]{neutron star\nobreakdashes--black hole}
\acro{NSBH}[NSBH]{\acl{NS}\nobreakdashes--\acl{BH}}
\acro{NSF}[NSF]{National Science Foundation}
\acro{NSNS}[NSNS]{\acl{NS}\nobreakdashes--\acl{NS}}
\acro{NS}[NS]{neutron star}
\acro{O1}[O1]{\acl{aLIGO}'s first observing run}
\acro{O2}[O2]{\acl{aLIGO}'s second observing run}
\acro{O3}[O3]{\acl{aLIGO}'s and \acl{AdVirgo} third observing run}
\acro{oLIB}[\acrolowercase{o}LIB]{Omicron+\acl{LIB}}
\acro{OT}[OT]{optical transient}
\acro{P48}[P48]{Palomar 48~inch Oschin telescope}
\acro{P60}[P60]{robotic Palomar 60~inch telescope}
\acro{P200}[P200]{Palomar 200~inch Hale telescope}
\acro{PC}[PC]{photon counting}
\acro{PESSTO}[PESSTO]{Public ESO Spectroscopic Survey of Transient Objects}
\acro{PSD}[PSD]{power spectral density}
\acro{PSF}[PSF]{point-spread function}
\acro{PS1}[PS1]{Pan\nobreakdashes-STARRS~1}
\acro{PTF}[PTF]{Palomar Transient Factory}
\acro{QUEST}[QUEST]{Quasar Equatorial Survey Team}
\acro{RAPTOR}[RAPTOR]{Rapid Telescopes for Optical Response}
\acro{REU}[REU]{Research Experiences for Undergraduates}
\acro{RMS}[RMS]{root mean square}
\acro{ROTSE}[ROTSE]{Robotic Optical Transient Search}
\acro{S5}[S5]{\ac{LIGO}'s fifth science run}
\acro{S6}[S6]{\ac{LIGO}'s sixth science run}
\acro{SAA}[SAA]{South Atlantic Anomaly}
\acro{SHB}[SHB]{short, hard burst}
\acro{SHGRB}[SHGRB]{short, hard \acl{GRB}}
\acro{SKA}[SKA]{Square Kilometer Array}
\acro{SMT}[SMT]{Slewing Mirror Telescope\acroextra{ (instrument on \acs{UFFO} Pathfinder)}}
\acro{SNR}[S/N]{signal\nobreakdashes-to\nobreakdashes-noise ratio}
\acro{SEDM}[SEDM]{Spectral Energy Distribution Machine}
\acro{SSC}[SSC]{synchrotron self\nobreakdashes-Compton}
\acro{SDSS}[SDSS]{Sloan Digital Sky Survey}
\acro{SED}[SED]{spectral energy distribution}
\acro{SFR}[SFR]{star formation rate}
\acro{SGRB}[SGRB]{short \acl{GRB}}
\acro{SN}[SN]{supernova}
\acroplural{SN}[SN\acrolowercase{e}]{supernova}
\acro{SNIa}[\acs{SN}\,I\acrolowercase{a}]{Type~Ia \ac{SN}}
\acroplural{SNIa}[\acsp{SN}\,I\acrolowercase{a}]{Type~Ic \acp{SN}}
\acro{SNIcBL}[\acs{SN}\,I\acrolowercase{c}\nobreakdashes-BL]{broad\nobreakdashes-line Type~Ic \ac{SN}}
\acroplural{SNIcBL}[\acsp{SN}\,I\acrolowercase{c}\nobreakdashes-BL]{broad\nobreakdashes-line Type~Ic \acp{SN}}
\acro{SVD}[SVD]{singular value decomposition}
\acro{TAROT}[TAROT]{T\'{e}lescopes \`{a} Action Rapide pour les Objets Transitoires}
\acro{TDOA}[TDOA]{time delay on arrival}
\acroplural{TDOA}[TDOA\acrolowercase{s}]{time delays on arrival}
\acro{TD}[TD]{time domain}
\acro{TOA}[TOA]{time of arrival}
\acroplural{TOA}[TOA\acrolowercase{s}]{times of arrival}
\acro{TOO}[ToO]{target\nobreakdashes-of\nobreakdashes-opportunity}
\acroplural{TOO}[ToO\acrolowercase{s}]{targets of opportunity}
\acro{TNS}[TNS]{Transient Name Server}
\acro{UFFO}[UFFO]{Ultra Fast Flash Observatory}
\acro{UHE}[UHE]{ultra high energy}
\acro{UVOT}[UVOT]{UV/Optical Telescope\acroextra{ (instrument on \emph{Swift})}}
\acro{VHE}[VHE]{very high energy}
\acro{VISTA}[VISTA@ESO]{Visible and Infrared Survey Telescope}
\acro{VLA}[VLA]{Karl G. Jansky Very Large Array}
\acro{VLT}[VLT]{Very Large Telescope}
\acro{VST}[VST@ESO]{\acs{VLT} Survey Telescope}
\acro{WAM}[WAM]{Wide\nobreakdashes-band All\nobreakdashes-sky Monitor\acroextra{ (instrument on \emph{Suzaku})}}
\acro{WASP}[WASP]{Wafer-Scale Imager for Prime}
\acro{WCS}[WCS]{World Coordinate System}
\acro{WISE}[WISE]{Wide-field Infrared Survey Explorer}
\acro{WSS}[w.s.s.]{wide\nobreakdashes-sense stationary}
\acro{XRF}[XRF]{X\nobreakdashes-ray flash}
\acroplural{XRF}[XRF\acrolowercase{s}]{X\nobreakdashes-ray flashes}
\acro{XRT}[XRT]{X\nobreakdashes-ray Telescope\acroextra{ (instrument on \emph{Swift})}}
\acro{ZTF}[ZTF]{Zwicky Transient Facility}
\end{acronym}